\documentclass{aa}  

\usepackage{graphicx}
\usepackage{txfonts}
\RequirePackage[normalem]{ulem} %DIF PREAMBLE
\RequirePackage{color}\definecolor{RED}{rgb}{1,0,0}\definecolor{BLUE}{rgb}{0,0,1} %DIF PREAMBLE
\providecommand{\DIFaddbegin}{} %DIF PREAMBLE
\providecommand{\DIFaddend}{} %DIF PREAMBLE
\providecommand{\DIFdelbegin}{} %DIF PREAMBLE
\providecommand{\DIFdelend}{} %DIF PREAMBLE
\providecommand{\DIFaddbeginFL}{} %DIF PREAMBLE
\providecommand{\DIFaddendFL}{} %DIF PREAMBLE
\providecommand{\DIFdelbeginFL}{} %DIF PREAMBLE
\providecommand{\DIFdelendFL}{} %DIF PREAMBLE
\newcommand{\DIFscaledelfig}{0.5}
\RequirePackage{settobox} %DIF PREAMBLE
\RequirePackage{letltxmacro} %DIF PREAMBLE
\newsavebox{\DIFdelgraphicsbox} %DIF PREAMBLE
\newlength{\DIFdelgraphicswidth} %DIF PREAMBLE
\newlength{\DIFdelgraphicsheight} %DIF PREAMBLE
\LetLtxMacro{\DIFOincludegraphics}{\includegraphics} %DIF PREAMBLE
\newcommand{\DIFaddincludegraphics}[2][]{{\color{blue}\fbox{\DIFOincludegraphics[#1]{#2}}}} %DIF PREAMBLE
\newcommand{\DIFdelincludegraphics}[2][]{% %DIF PREAMBLE
        \sbox{\DIFdelgraphicsbox}{\DIFOincludegraphics[#1]{#2}}% %DIF PREAMBLE
        \settoboxwidth{\DIFdelgraphicswidth}{\DIFdelgraphicsbox} %DIF PREAMBLE
        \settoboxtotalheight{\DIFdelgraphicsheight}{\DIFdelgraphicsbox} %DIF PREAMBLE
        \scalebox{\DIFscaledelfig}{% %DIF PREAMBLE
                \parbox[b]{\DIFdelgraphicswidth}{\usebox{\DIFdelgraphicsbox}\\[-\baselineskip] \rule{\DIFdelgraphicswidth}{0em}}\llap{\resizebox{\DIFdelgraphicswidth}{\DIFdelgraphicsheight}{% %DIF PREAMBLE
                                \setlength{\unitlength}{\DIFdelgraphicswidth}% %DIF PREAMBLE
                                \begin{picture}(1,1)% %DIF PREAMBLE
                                \thicklines\linethickness{2pt} %DIF PREAMBLE
                                {\color[rgb]{1,0,0}\put(0,0){\framebox(1,1){}}}% %DIF PREAMBLE
                                {\color[rgb]{1,0,0}\put(0,0){\line( 1,1){1}}}% %DIF PREAMBLE
                                {\color[rgb]{1,0,0}\put(0,1){\line(1,-1){1}}}% %DIF PREAMBLE
                                \end{picture}% %DIF PREAMBLE
                        }\hspace*{3pt}}} %DIF PREAMBLE
} %DIF PREAMBLE
\LetLtxMacro{\DIFOaddbegin}{\DIFaddbegin} %DIF PREAMBLE
\LetLtxMacro{\DIFOaddend}{\DIFaddend} %DIF PREAMBLE
\LetLtxMacro{\DIFOdelbegin}{\DIFdelbegin} %DIF PREAMBLE
\LetLtxMacro{\DIFOdelend}{\DIFdelend} %DIF PREAMBLE
\DeclareRobustCommand{\DIFaddbegin}{\DIFOaddbegin \let\includegraphics\DIFaddincludegraphics} %DIF PREAMBLE
\DeclareRobustCommand{\DIFaddend}{\DIFOaddend \let\includegraphics\DIFOincludegraphics} %DIF PREAMBLE
\DeclareRobustCommand{\DIFdelbegin}{\DIFOdelbegin \let\includegraphics\DIFdelincludegraphics} %DIF PREAMBLE
\DeclareRobustCommand{\DIFdelend}{\DIFOaddend \let\includegraphics\DIFOincludegraphics} %DIF PREAMBLE
\LetLtxMacro{\DIFOaddbeginFL}{\DIFaddbeginFL} %DIF PREAMBLE
\LetLtxMacro{\DIFOaddendFL}{\DIFaddendFL} %DIF PREAMBLE
\LetLtxMacro{\DIFOdelbeginFL}{\DIFdelbeginFL} %DIF PREAMBLE
\LetLtxMacro{\DIFOdelendFL}{\DIFdelendFL} %DIF PREAMBLE
\DeclareRobustCommand{\DIFaddbeginFL}{\DIFOaddbeginFL \let\includegraphics\DIFaddincludegraphics} %DIF PREAMBLE
\DeclareRobustCommand{\DIFaddendFL}{\DIFOaddendFL \let\includegraphics\DIFOincludegraphics} %DIF PREAMBLE
\DeclareRobustCommand{\DIFdelbeginFL}{\DIFOdelbeginFL \let\includegraphics\DIFdelincludegraphics} %DIF PREAMBLE
\DeclareRobustCommand{\DIFdelendFL}{\DIFOaddendFL \let\includegraphics\DIFOincludegraphics} %DIF PREAMBLE
\begin{document}

   \title{The observational evidence that all microflares that accelerate electrons to high energies are rooted in sunspots}
   \titlerunning{All microflares that accelerate electrons to high-energies are rooted in sunspots}

   \author{Andrea Francesco Battaglia \inst{1,2,3}
          \and
          Säm Krucker \inst{2,4}
          \and 
          Astrid M. Veronig \inst{5,6}
          \and
          Muriel Zoë Stiefel \inst{2,3}
          \and
          \\
          Alexander Warmuth \inst{7}
          \and
          Arnold O. Benz \inst{2,3}
          \and
          Daniel F. Ryan \inst{2}
          \and
          Hannah Collier \inst{2,3}
          \and
          Louise Harra \inst{3,8}
          }

   \institute{
        Istituto ricerche solari Aldo e Cele Daccò (IRSOL), Faculty of informatics, Università della Svizzera italiana, Locarno, Switzerland \\
        \email{andrea.francesco.battaglia@irsol.usi.ch}
        \and
        University of Applied Sciences and Arts Northwestern Switzerland (FHNW), Bahnhofstrasse 6, 5210 Windisch, Switzerland
        \and
        Institute for Particle Physics and Astrophysics (IPA), Swiss Federal Institute of Technology in Zurich (ETHZ), Wolfgang-Pauli-Strasse 27, 8039 Zurich, Switzerland
        \and
        Space Sciences Laboratory, University of California, 7 Gauss Way, 94720 Berkeley, USA
        \and 
        Institute of Physics, University of Graz, Universit\"atsplatz 5, A-8010 Graz, Austria
        \and
        Kanzelh\"ohe Observatory for Solar and Environmental Research, University of Graz, Kanzelh\"ohe 19, 9521 Treffen, Austria
        \and
        Leibniz-Institut f\"ur Astrophysik Potsdam (AIP), An der Sternwarte 16, D-14482 Potsdam, Germany
        \and
        Physikalisch-Meteorologisches Observatorium Davos, World Radiation Center, 7260 Davos Dorf, Switzerland 
             }

   \date{Received XX month 2024 / Accepted YY month 2024}

% \abstract{}{}{}{}{} 
% 5 {} token are mandatory
 
  \abstract
   {In general, large solar flares are more efficient at accelerating high-energy electrons than microflares. Nonetheless, sometimes microflares that accelerate electrons to high energies are observed. Their origin is unclear.}
   {We statistically characterized microflares with strikingly hard spectra in the hard X-ray (HXR) range, which means that they are efficient at accelerating high-energy electrons. We refer to these events as "hard microflares."}
   {We selected 39 hard microflares, based on their spectral hardness estimated from the Solar Orbiter/STIX HXR quicklook light curves in two energy bands. The statistical analysis is built on spectral and imaging information from STIX combined with extreme ultraviolet (EUV) and magnetic field maps from SDO/AIA and SDO/HMI.}
   {The key observational result is that all hard microflares in this dataset have one of the footpoint rooted directly within a sunspot (either in the umbra or the penumbra). This clearly indicates that the underlying magnetic flux densities are large. For the events with the classic two-footpoints morphology, the absolute value of the mean line-of-sight magnetic flux density (and vector magnetic field strength) at the footpoint rooted within the sunspot ranges from 600 to 1800 G (1500 to 2500 G), whereas the outer footpoint measures from 10 to 200 G (100 to 400 G), therefore about ten times weaker.
   In addition, approximately 78\% of the hard microflares, which exhibited two HXR footpoints, have similar or even stronger HXR flux from the footpoint rooted within the sunspot. This contradicts the magnetic mirroring scenario. The median footpoint separation, measured through HXR observations, is approximately 24 Mm, which aligns with regular events of similar GOES classes. In addition, about 74\% of the events could be approximated by a single-loop geometry, demonstrating that hard microflares typically have a relatively simple morphology. Out of these events, around 54\% exhibit a relatively flat flare loop geometry.
   }
   {We conclude that all hard microflares are rooted in sunspots, which implies that the magnetic field strength plays a key role in efficiently accelerating high-energy electrons, with hard HXR spectra associated with strong fields. This key result will allow us to further constrain our understanding of the electron acceleration mechanisms in flares and space plasmas.}


   \maketitle
%
%-------------------------------------------------------------------

\section{Introduction}

To understand the physics underpinning solar flares, it is crucial to understand how particles are accelerated to high energies. Accelerated particles carry a significant amount of the total flare energy \citep[e.g.,][]{1995ARA&A..33..239H}.

In the context of solar flares, three primary particle acceleration mechanisms have been reported in the literature \citep{2004psci.book.....A}: DC electric field acceleration \citep[e.g.,][]{1985ApJ...293..584H,2006Natur.443..553D}, stochastic acceleration \citep[e.g.,][]{1992wapl.book.....S}, and shock acceleration \citep[e.g.,][]{1985ApJ...298..400E}. Electric field acceleration, proposes the acceleration within quasi-steady electric fields, which can be produced in current sheets, during magnetic reconnection events, or within current-carrying loops. Stochastic acceleration mechanisms revolve around random energy gains and losses in a turbulent plasma \citep{2004psci.book.....A}. Shock acceleration mechanisms involve an inhomogeneous shock front that is suitable to transfer momentum and energy to intercepted particles \citep{2004psci.book.....A}.

Regardless of the specific mechanism, particles are accelerated in the corona. These accelerated particles either move outward along "open" field lines to escape to interplanetary space or downward along closed field lines towards the denser chromosphere, where they emit radiation across the entire electromagnetic spectrum, including hard X-rays (HXRs). Although HXRs account for only a small fraction of the total radiated energy, they can serve as a useful diagnostic tool, as they allow one to infer the energy of the accelerated particles deposited in the chromosphere \citep{2011SSRv..159...19F,2017LRSP...14....2B}. %, under the assumption of the collisional thick target model \citep{1971SoPh...18..489B}. 
In this study, we focus on the energy range from 4 to 150 keV, which are photons mainly produced by electrons.

The accelerated electrons moving downward deposit their energy into the chromosphere and lower transition region, heating the plasma to high temperatures observable in the soft X-ray (SXR) range \citep{2017LRSP...14....2B}. This heated plasma then follows the flare loop, extending into the corona via a process known as "chromospheric evaporation" \citep[e.g.,][]{1974SoPh...34..323H,1984ApJ...287..917A,1985ApJ...289..414F,2002A&A...392..699V}. The observable correlation between the SXR flux originating from the heated plasma and the cumulative HXR flux from the accelerated electrons is known as the Neupert effect \citep{1968ApJ...153L..59N}.

Evaporation can be categorized into two types: explosive \citep[e.g.,][]{2006ApJ...638L.117M} and gentle \citep[e.g.,][]{2006ApJ...642L.169M}. These categories are distinguished by the speed at which the chromospheric plasma, once heated, expands into the corona. Studies indicate that the velocity of this expansion depends on the intensity of the accelerated electron beam that reaches the chromospheric plasma \citep[e.g.,][]{1985ApJ...289..414F}. Moreover, the electron spectrum plays a crucial role, with low-energy electrons proving more effective at heating the atmosphere (which subsequently triggers evaporation) compared to their high-energy counterparts \citep{2015ApJ...808..177R}.

The amount of chromospheric evaporation is related to the plasma heating and the emitted SXR flux. This SXR emission is a defining factor in differentiating a (standard) flare from a microflare.\footnote{We refer to Appendix~\ref{sec:event-selection} for the description of what we consider to be a microflare.} Microflares are dynamic, small-scale energy release events with energy several orders of magnitude smaller than regular flares. Despite their relatively smaller scale, microflares exhibit intriguing characteristics. For instance, standard (or regular) microflares are known for their steep (or soft) HXR spectra, which suggests that they are less efficient at accelerating high-energy electrons compared to larger flares \cite[e.g.,][]{2005A&A...439..737B,2008ApJ...677..704H,2011SSRv..159..263H,2014ApJ...789..116I,2016A&A...588A.116W}. However, there are microflare observations of remarkably hard spectra \citep[e.g., ][]{2008A&A...481L..45H,2013ApJ...765..143I,2018ApJ...856..111L,2023A&A...670A..56B,2024A&A...683A..41S}, which implies the presence of prominent nonthermal emission due to high energy electrons. This indicates that factors other than the flare energy significantly influence the acceleration efficiency. Investigating such factors would allow for further constraining of the acceleration mechanism in flares. 

In a recent case study of two microflares with hard spectra, \citet{2024A&A...683A..41S} found that both events have one of their footpoints directly rooted within a sunspot (either in the umbra or the penumbra), suggesting that the strong magnetic field of the sunspot may be responsible for the hard microflare spectrum.
From observations taken by the Spectrometer/Telescope for Imaging X-rays \citep[STIX;][]{2020A&A...642A..15K}, the X-ray telescope on board the Solar Orbiter mission \citep{2020A&A...642A...1M}, we collected 39 microflares characterized by hard HXR spectra, which we call "hard microflares." %Our statistical analysis included HXR images with UV and visible images. 
The location of the footpoints in relation to the associated sunspot was determined. %We used these events to infer additional properties, in order to address open questions regarding the acceleration mechanism and the subsequent thermal response. 
%We also discuss hard microflares in the context of previous studies in the literature of similar events. 

Section~\ref{sec:data-analysis} details the data analysis. The results are presented in Sect.~\ref{sec:results}. In Sect.~\ref{sec:discussions} we discuss our findings. Finally, our conclusions are drawn in Sect.~\ref{sec:conslusions}.

%--------------------------------------------------------------------

\section{Event selection and data analysis \label{sec:data-analysis}}

In this section we describe the data analysis of STIX, AIA and HMI observations. In our analysis, we examined 39 hard microflares, as detailed in Table~\ref{tab:all-hard-microflares}. The selection method for these microflares is described in Appendix~\ref{sec:event-selection}.

%----------

    \subsection{STIX data analysis}

    In order to study the X-ray signal of the selected hard microflares, we use the observations by STIX, the X-ray telescope aboard Solar Orbiter. Due to the varying distance between the Solar Orbiter spacecraft and the Sun compared to the Earth, there is a difference in the photon arrival time. To account for this difference, all time measurements made by STIX have been adjusted and are expressed in Earth UT.

    As hard microflares tend to have a relatively short high-energy burst (tens of seconds to a few minutes), as shown in Fig.~\ref{fig:example-quicklook}, we use the STIX quicklook data for event selection (see Appendix~\ref{sec:event-selection}). This data product aids automatic detection for two reasons: firstly, this data is continuously accessible at a 4 s cadence, unlike spectrogram or pixel data where the integration time changes throughout each flare, and secondly, the same energy bands are consistently available.
    The STIX ground software (version 0.5.2, as of March 2024) was used for the preparation of the STIX data.

    \begin{figure*}[!]
        \centering
        \includegraphics[width=0.95\textwidth]{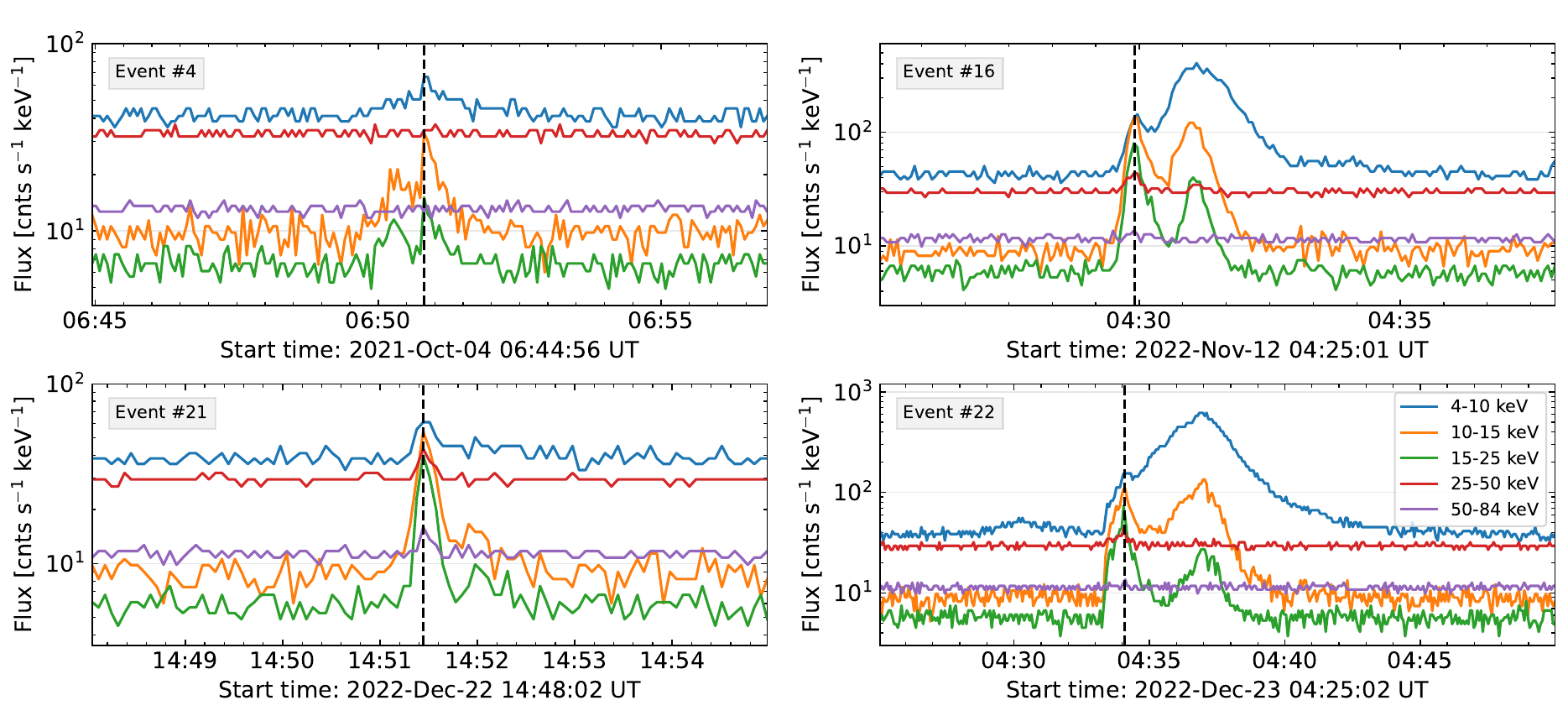}
        \caption{STIX quicklook time profiles of four representative hard microflares from our sample. The energy range of the different curves is indicated in the legend located in the bottom right panel. The background-subtracted GOES classes for these events are A2 (event \#4), B4 (event \#16), A8 (event \#21), and C1 (event \#22), respectively. The vertical black dashed line represents the nonthermal peak time.}
        \label{fig:example-quicklook}
    \end{figure*}

    \begin{figure*}
            \centering
            \includegraphics[width=0.95\textwidth]{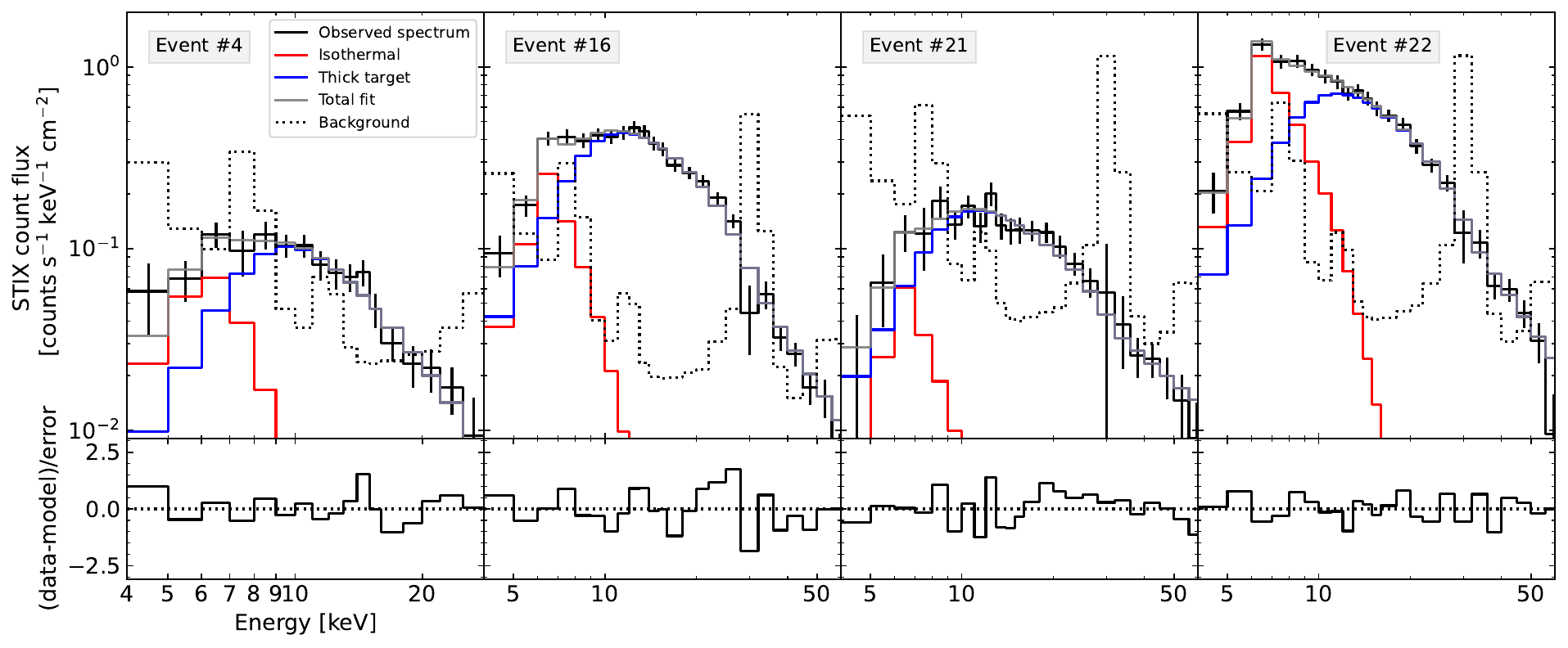}
            \caption{STIX spectra for the same events shown in Fig.~\ref{fig:example-quicklook}. The solid black line illustrates the observed, background-subtracted flare spectra, while the dashed black lines display the background. The thermal and nonthermal models are represented by the red and blue curves, respectively. The sum of the two models is shown as a gray line. Beneath each spectrum, the residuals are provided, which are calculated by subtracting the total fit from the data and normalizing the result by the error bar.}
            \label{fig:example-spectra}
        \end{figure*}
    
    %----------
    
        \subsubsection{X-ray spectroscopy}
    
        For the spectral fitting of the STIX observations, spectrogram data have been used. The spectral fitting has been performed using the Object Spectral Executive \citep[OSPEX;][]{2002SoPh..210..165S} software available through the SolarSoftWare \citep[SSW;][]{1998SoPh..182..497F}.

        In this work, all spectral parameters were derived from fitting the STIX spectra around the time of the peak at high energies. The integration time around this peak is typically 10 to 20 seconds, depending on the counting statistics. %, usually corresponding to one or two time bins. This is due to the short duration of the bursts and the relatively low counting statistics of these events combined with the dynamic time binning feature of the STIX flight software.

        In Fig.~\ref{fig:example-spectra}, we present four different STIX spectra along with the fits as examples. To obtain the background-subtracted spectra, we only subtracted the STIX observations taken during quiet times, closest to the flare events. This is because, during quiet times, the signal is dominated by the onboard calibration source. The low-energy part of the spectrum was fitted with a standard isothermal model (the \texttt{vth} function in OSPEX). Here, we used coronal abundances, with the default set in the CHIANTI database version 10.0.1 \citep{1997A&AS..125..149D,2020Atoms...8...46D}.
        
        Due to the nature of the high-energy peaks, which are impulsive, relatively short and decrease rapidly, they have been interpreted as nonthermal emission originating from accelerated electrons injected into the chromosphere. This interpretation is also supported by X-ray and ultraviolet (UV) images (Sect.~\ref{sec:results}). Therefore, to fit the high-energy part of the spectrum, we used the standard thick-target model \texttt{thick2\_vnorm} function available in OSPEX. For this fitting, we considered only a single power-law, and the resulting low-energy cutoff is always around 10 to 20 keV.

        %In summary, all spectral parameters shown in this paper are derived using a combination of thermal and nonthermal models around the high-energy peak, which corresponds to the peak of the nonthermal emission.

    %----------

        \subsubsection{X-ray imaging}    

        In the reconstruction of the STIX images, similar to the spectroscopy analysis, we considered times around the nonthermal peak (i.e., the high-energy peak). Given that the events discussed in this paper exhibit hard spectra, and that the STIX background between 10 to 20 keV is the lowest, which corresponds also to the energy range where microflares typically exhibit nonthermal emission, this allows for easier nonthermal image reconstruction compared to thermal images. This is not typically the case, since for standard microflares, if imaging reconstruction can be done, this is mostly in the thermal range. An illustrative example is the second event from the left in Fig.~\ref{fig:example-spectra}. Here, we have a high count rate to use for nonthermal image reconstruction, whereas the thermal component is below the background rate. In such a case, it is not possible to obtain reliable image reconstructions of the thermal microflare emission. When this situation occurs, we considered times around the thermal peak for obtaining a thermal image. However, for many events, the thermal and nonthermal peak roughly coincide, making it impossible to obtain a reliable thermal image.
        In this paper, we considered 39 events (refer to Sect.~\ref{sec:event-selection}): Thermal images could be obtained from 28 events and nonthermal images from almost all events, 34 in total.
        
        The CLEAN algorithm \citep{1974A&AS...15..417H} has been used to reconstruct all images, using natural weighting. The size of the selected CLEAN beam corresponds to the angular resolution of the sub-collimators associated with grid label 3 \citep{2020A&A...642A..15K}, which is 14.6 arcsec. The sub-collimators associated with the finest grids, labeled 1 and 2, were excluded from the analysis as their calibration is still in progress.

        For image co-alignment, the nonthermal images were aligned with the UV sources observed in the AIA 1600 \AA{} maps, which typically show the flare ribbons. More details are provided in the following subsection.
        
        To estimate the X-ray flux coming from each nonthermal source, we used the STIX visibility forward-fit algorithm \citep{2022A&A...668A.145V}. We assumed that the sources were circular Gaussians, as they tend to be compact sources in hard microflares, particularly when located within the sunspots. The location of each source was then fixed, based on the locations obtained from the previously generated CLEAN image. Consequently, while the location was kept fixed, both the sizes of the sources and their respective fluxes were allowed to vary.

%----------

    \subsection{AIA and HMI data analysis \label{subsec:AIAandHMI}}

    The data from the Atmospheric Imaging Assembly \citep[AIA;][]{2012SoPh..275...17L} and the Helioseismic and Magnetic Imager \citep[HMI;][]{2012SoPh..275..207S}, instruments aboard the Solar Dynamics Observatory \citep[SDO;][]{2012SoPh..275....3P}, were obtained from the Joint Science Operations Center (JSOC). The AIA maps were calibrated using the aiapy software \citep{2020JOSS....5.2801B}. Visualizations, including images, histograms, spectra, scatter plots, and time profiles, were generated using SunPy \citep{2020ApJ...890...68S}.

    To overlay the STIX reconstructed images on top of the AIA maps, we first reprojected the AIA 1600 \AA{} images to match the Solar Orbiter vantage point using the standard tools provided by SunPy. We then co-aligned the STIX images with the reprojected maps, by aligning the nonthermal sources to align with the flare ribbons that can be observed in the UV maps.

    In order to estimate the photospheric magnetic field density at the footpoints of the flare loops, we used the line-of-sight (LoS) magnetic field data from the HMI magnetograms (in Appendix~\ref{sec:vec-magn-field}, we used vector magnetic field data). We obtained the flux density by averaging over the flare ribbon area, as defined by the AIA 1600 \AA{} images. The uncertainties on the magnetic field density derives from the standard deviation within the same flare ribbon area. For this part of the analysis, we only considered events located within 50 deg longitude from the disk center as seen from Earth, which are 27 events in total.

%--------------------------------------------------------------------

\section{Results \label{sec:results}}

%\textbf{Rephrase.} In this section we present the results obtained from the observations in visible, UV and X-rays of hard microflares. We first of all looked into the images of these events, then analyzed them statistically by considering HXR parameters together with the LoS magnetic field density. We also present the method used to automatically get the list of events analyzed in this paper.

    %----------

    \subsection{Visible, UV, EUV, and HXR images}

    Since all the STIX events discussed in this paper were also observed from Earth, we compared the location of the flare ribbons with the location of the source active regions and sunspots using SDO imaging in the UV and visible. Afterwards, we investigated the flare morphology by comparing the AIA and HMI observations with the STIX images.

        %----------

        \subsubsection{Footpoint locations}
        
\begin{figure*}
            \centering
            \includegraphics[width=\textwidth]{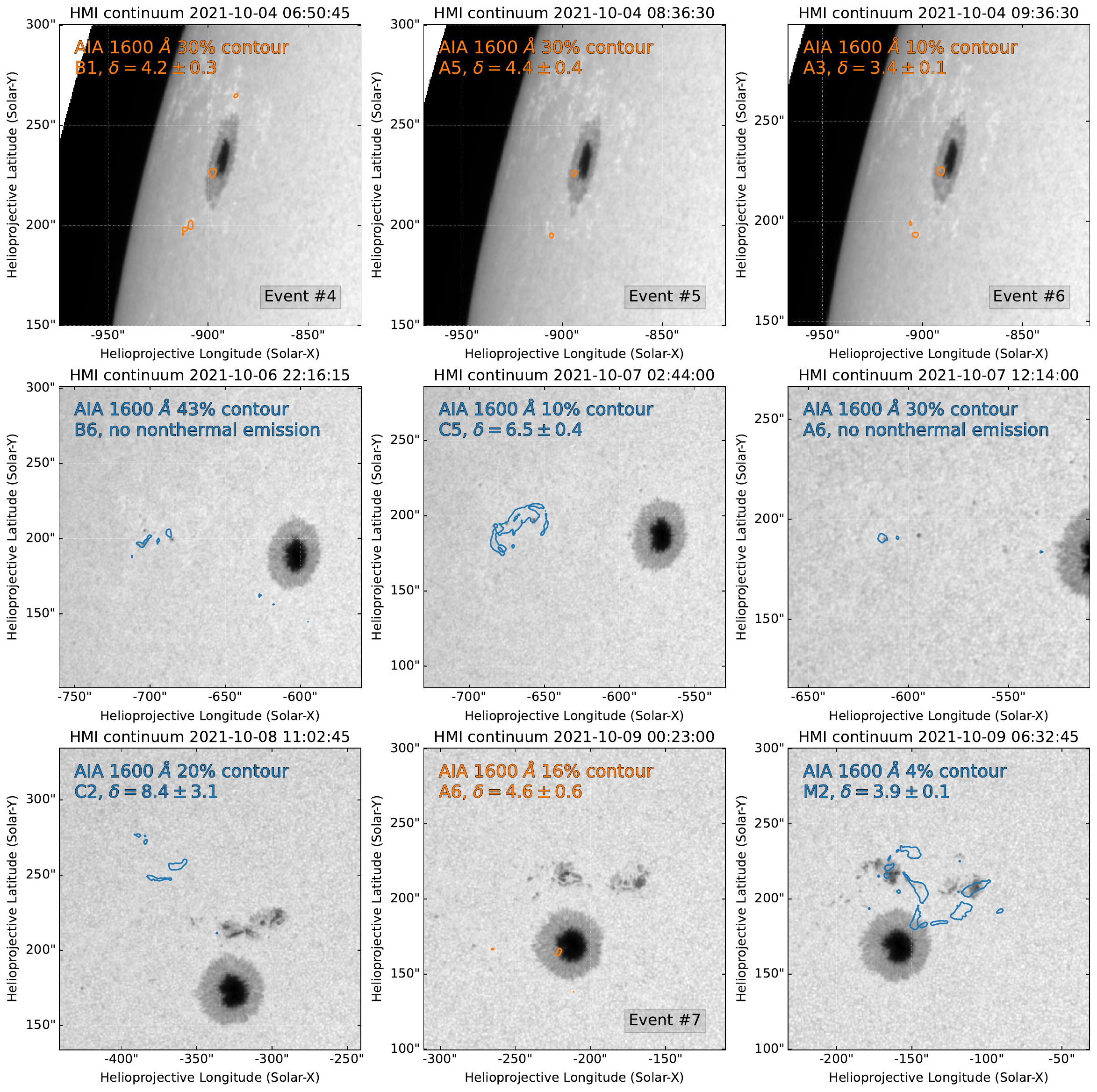}
            \caption{Flare ribbon location of various types of flares (standard microflares, hard microflares, and a medium-size flare) that occurred within AR12882. 
            The orange contours represent the flare ribbons identified in the AIA 1600 \AA{} images of the hard microflares, while the blue contours represent all other events.
            The intensity map from SDO/HMI is plotted in the background. The legend includes the percentage of the AIA 1600 \AA{} contour used, the GOES class, and the electron spectral index $\delta$. This figure clearly shows that hard microflares are rooted in sunspots (umbra or penumbra), while standard microflares, which have a soft spectrum or a total lack of nonthermal emission, are located far from sunspots.}
            \label{fig:location-hardVSsoft}
        \end{figure*}

        Before examining the location of all hard microflares, we focused on one active region, AR12882. Our aim was to compare the location of different types of flares, such as "standard" microflares (those with softer spectra or even no nonthermal emission), hard microflares, and relatively large flares. Figure~\ref{fig:location-hardVSsoft} displays some events that occurred within AR12882 from October 4th, 2021, to October 10th, 2021. This figure clearly shows that hard microflares are rooted directly in sunspots, whereas standard microflares are located away from the sunspots in the plage regions surrounding the AR. This is in agreement with what is reported in \citet{2024A&A...683A..41S}. Larger flares, such as the M2 class flare in Fig.~\ref{fig:location-hardVSsoft}, are more spatially extended and eventually cross sunspot areas during their time evolution. This is already known and reported in the literature \citep[e.g.,][]{2012ApJ...747..134M,2016ApJ...816...88K}.
    
        The novelty here is that all hard microflares are rooted in sunspots. This is not only true for the events from AR12882 but for all events listed in Tab.~\ref{tab:all-hard-microflares}. %This indicates that the hardness of the HXR spectrum is caused by the strength of the magnetic field in the sunspot area. This will be discussed in more details in Sect.~\ref{sec:discussions}.
    
\begin{figure*}
            \centering
            \includegraphics[width=\textwidth]{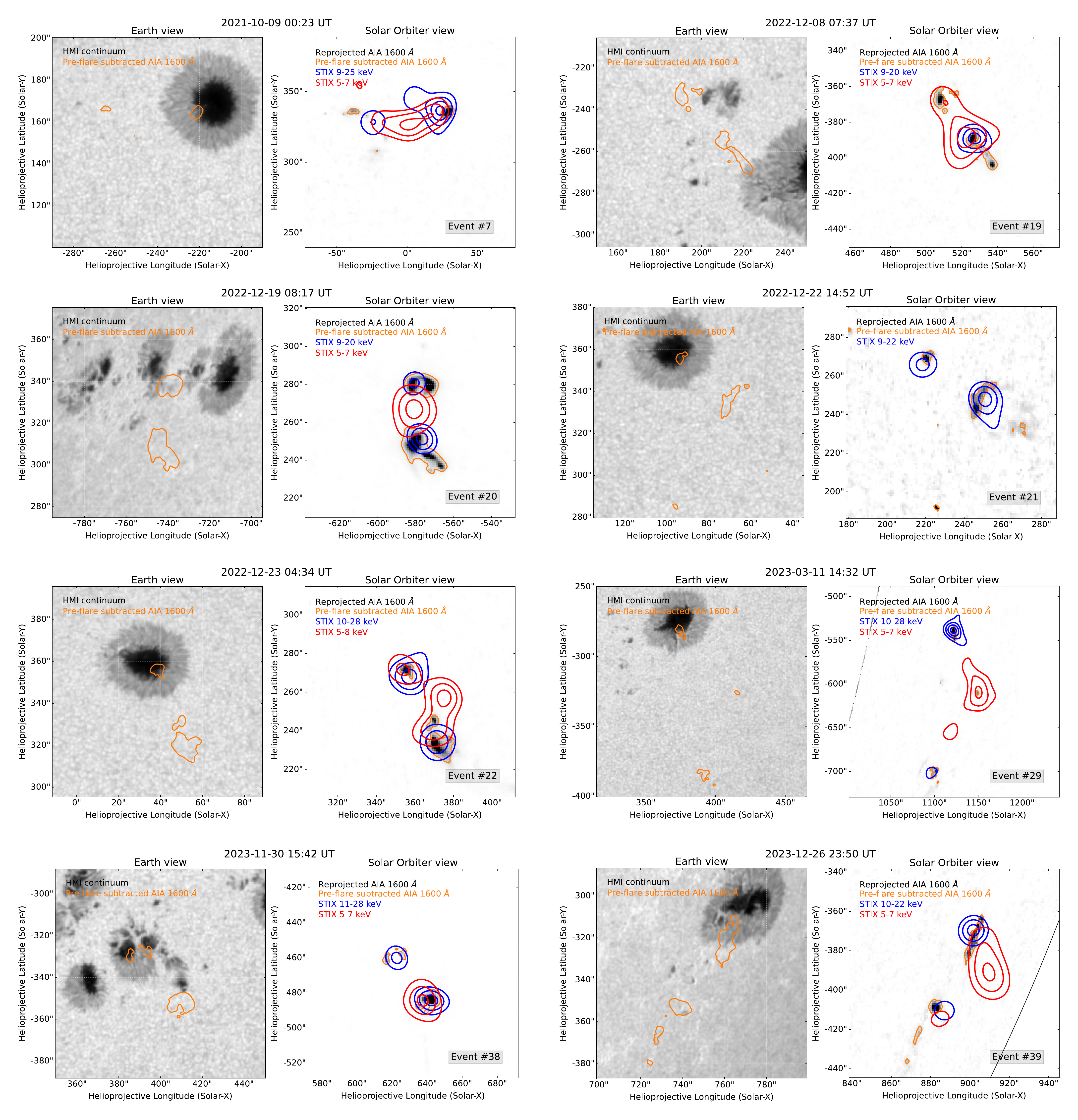}
            \caption{Solar Orbiter/STIX HXR images of eight hard microflares. For each event, the left panel displays the SDO/HMI intensitygram and the pre-flare subtracted SDO/AIA 1600 \AA{} contours (orange) from the Earth's perspective. The right panel shows the SDO/AIA 1600 \AA{} image closer to the nonthermal peak time reprojected to the Solar Orbiter view, with orange the reprojected pre-flare subtracted SDO/AIA 1600 \AA{} contours. The STIX images are displayed as red (thermal emission) and blue (nonthermal emission) contours. The event number is displayed in the bottom-right corner.}
            \label{fig:SDO-STIX-images}
        \end{figure*}

        It should be noted, however, that in some cases, part of one of the ribbons seem to be located at the edge of the penumbra, as it is the case of the event 20 in Fig.~\ref{fig:SDO-STIX-images}. For this reason, we introduced a category for distinguishing them from the events clearly rooted in sunspots. However, the fact that they appear to be at the edge of the sunspot, may be due to the inclination of the magnetic field lines, especially in the penumbra, as the location of the flare footpoints have been deduced from chromospheric altitudes. Therefore, an inclined magnetic field line in the photosphere, may result in a location at the edge of the sunspot if observed at chromospheric altitudes. 
        What is clear, is that the standard microflares are way more distant from sunspots than the hard microflares at the edge of sunspots. The standard microflares can be 50 arcsec away or more from the sunspots, while the hard microflares are within or at the edge. This indicates that they are still magnetically connected to the sunspot.
        It should be noted that the selection criteria of hard microflares impacts the number of detected events categorized as at the edge of sunspots (see Appendix~\ref{subsec:discussions-select-criterion}).
    
        %----------

        \subsubsection{Flare loop geometry \label{sec:flare-loop-geometry}}
        
\begin{figure*}[!]
            \centering
            \includegraphics[width=\textwidth]{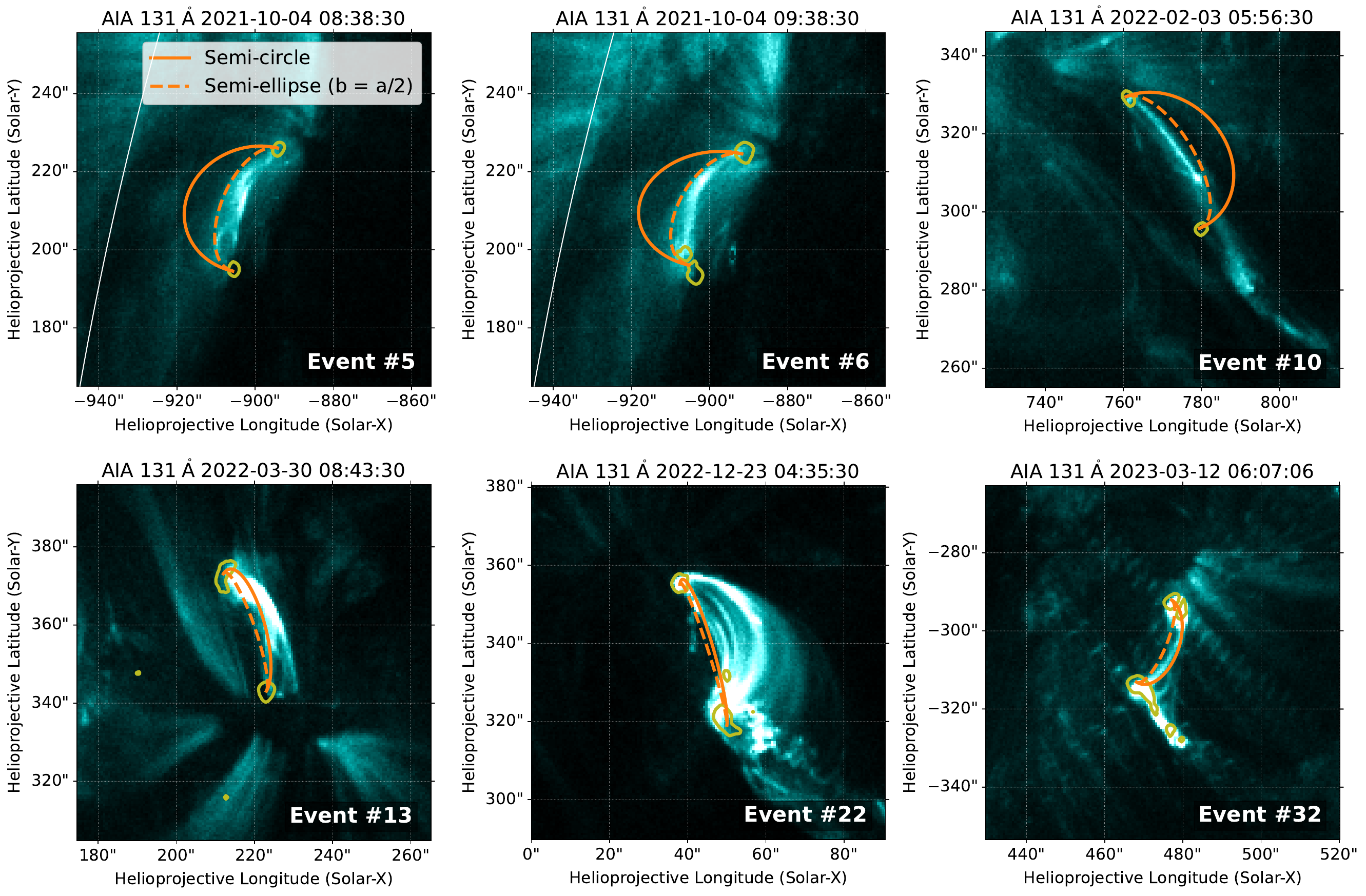}
            \caption{Heated flare loops of six hard microflares with SDO/AIA 131 \AA{} images. Overlaid on the AIA 131 \AA{} images are reference loops: a solid orange semi-circle perpendicular to the solar surface, with a radius defined as half the distance between the UV ribbons (AIA 1600 \AA{}, green contours), and a dashed orange semi-ellipse, where its semi-minor axis is half of the semi-major one. The latter is considered a reference for flat loops.}
            \label{fig:flares-with-loops}
        \end{figure*}

        Figure~\ref{fig:flares-with-loops} presents the AIA 131 \AA{} images of six hard microflare events. These images show the heated flare loop. To analyze their geometry, we overlaid two reference loops on top of the AIA images. One represents a standard semi-circle loop geometry (solid orange), and the other represents a flat geometry (dashed orange), which we defined as a semi-ellipse with its semi-minor axis being half of the semi-major axis. For these examples, events 5 and 6 appear to be flat loops. Both events originated from the same AR and similar location, suggesting a similar magnetic field morphology during their occurrence. Event 10 also seems to be a flat loop. This is not the case for events 13, 22, and 32. Event 22, however, seems to show a high degree of tilting.
        
        This type of analysis has been performed for all events in our sample (with the exception of event 12, as there is no AIA coverage) and their corresponding flare loop geometry can be found in the last column of Tab.~\ref{tab:all-hard-microflares}. If we could not determine whether a loop was flat or a semi-circle proxy, we labeled it either as "unclear" when it was difficult to distinguish between these two categories (for instance, during events close to the disk center), or as "complex" when the loop geometry could not be approximated by a single-loop, or as "tilted" when a significant degree of tilting was required.

        About 74\% (28 out of 38) of the events could be approximated by a simple loop geometry, demonstrating that hard microflares typically have a relatively simple geometry. Of these 28 events, it was not possible to distinguish between a semi-circle or flat loop geometry in four cases due to their orientation or location relative to the disk center. For the remaining 24 events, 13 ($54 \%$) displayed a flat geometry, eight ($33 \%$) had a semi-circle geometry, and three required a high degree of tilting.
        
        We note that projection effects may influence these results, as the representation of loops on a 2D image can be modified by changing the tilt or height of the loop. Therefore, to help the investigation of the loop geometry, magnetic field extrapolations could be beneficial.
        
%----------

    \subsection{HXR spectral parameters}

    We are now interested in comparing the electron spectral index and the photon flux derived from the thick target model of hard microflares with previous studies.

\begin{figure*}
        \centering
        \includegraphics[width=\textwidth]{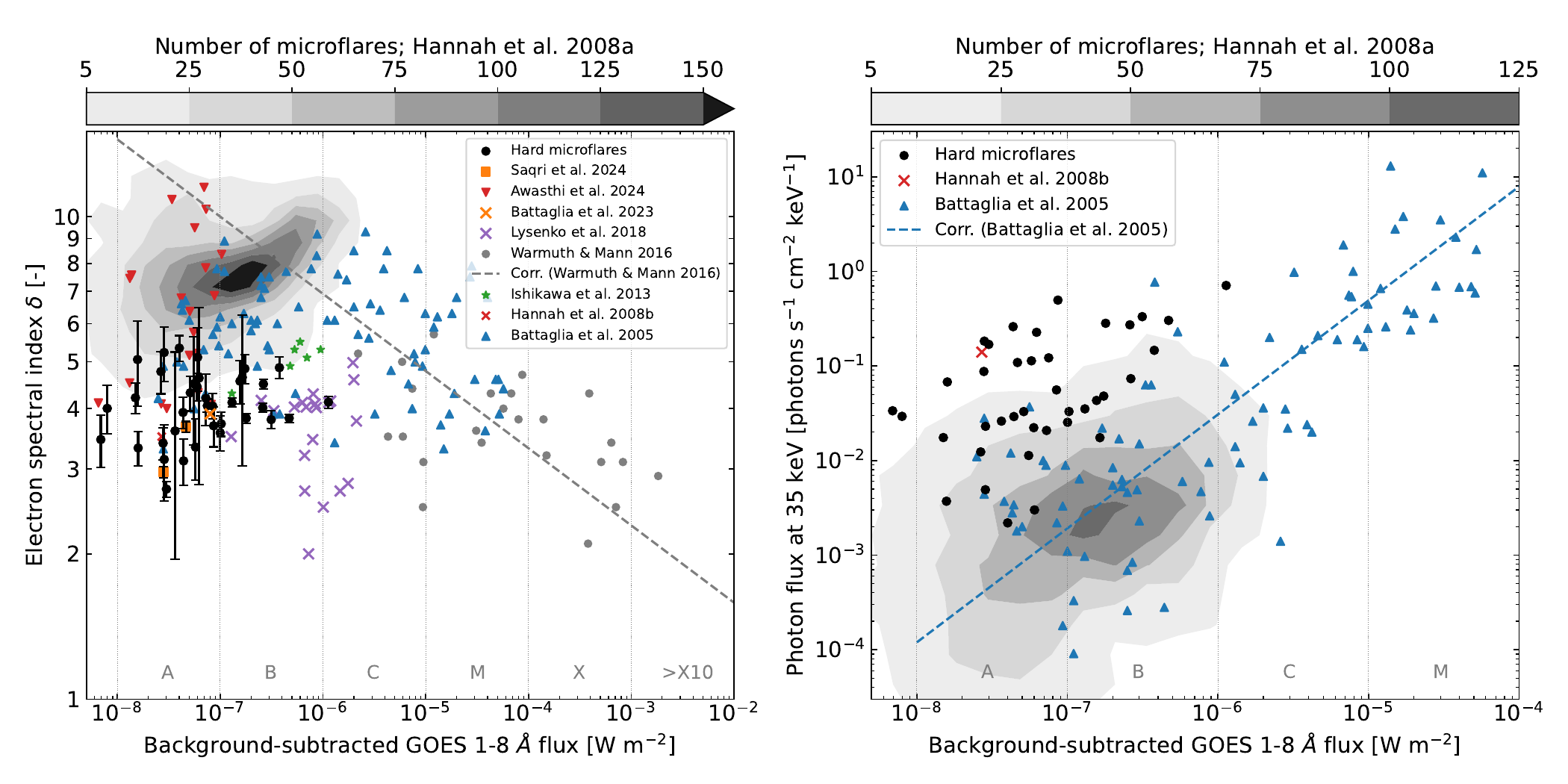}
        \caption{Comparison of the HXR spectral parameters of the hard microflares with events reported in the literature. (\emph{Left}) Electron spectral index $\delta$ as a function of the background-subtracted GOES 1-8 \AA{} SXR flux. Black circles represent the hard microflares considered in this paper. As a reference, we report studies of general flare and microflare samples \citep{2016A&A...588A.115W,2008ApJ...677..704H,2005A&A...439..737B}, and studies that included microflares with hard spectra \citep{2024A&A...683A..41S,2024ApJ...964..142A,2023A&A...670A..56B,2018ApJ...856..111L,2013ApJ...765..143I,2008A&A...481L..45H}. %while other symbols and colors refer to different studies already published \citep{}. 
        These studies are detailed in the legend. The gray contour levels correspond to the RHESSI microflare study by \citet{2008ApJ...677..704H}. The gray dashed line represents the correlation curve deduced by \citet{2016A&A...588A.115W} from the observations of flares above the GOES C1 level. (\emph{Right}) HXR photon flux at 35 keV plotted against the background-subtracted GOES 1-8 \AA{} SXR flux. Black dots represent the hard microflares. The blue triangles and the blue dashed curve are taken from \citet{2005A&A...439..737B}, while the red cross is from \citet{2008A&A...481L..45H}. The gray contour levels again refer to the RHESSI microflare study by \citet{2008ApJ...677..704H}.
        %\afbcomm{Correct the GOES factor for Marina, Ishikawa and Lysenko}
        }
        \label{fig:delta&ph35-GOES}
    \end{figure*}

    The left panel of Fig.~\ref{fig:delta&ph35-GOES} displays the electron spectral index against the background-subtracted GOES 1-8 \AA{} flux for different flares. This plot reveals several remarkable features. First of all, the correlation curve by \citet{2016A&A...588A.115W}, derived from flare observations of GOES classes larger than C1, aligns well with the independent studies involving microflares (GOES A and B class events) as reported in \citet{2008ApJ...677..704H} and \citet{2005A&A...439..737B}. This correlation suggests that the flare SXR intensity correlates with the hardness of its spectrum, which turns out to be true for both flare and microflare regimes.
    Secondly, previous studies on microflares with relatively hard spectra \citep{2024A&A...683A..41S,2024ApJ...964..142A,2023A&A...670A..56B,2018ApJ...856..111L,2013ApJ...765..143I,2008A&A...481L..45H} align well with the hard microflares analyzed in this paper.\footnote{It is important to note that the GOES flux values presented in Fig.~\ref{fig:delta&ph35-GOES} for the events discussed by \citet{2018ApJ...856..111L} differ from those reported in their original paper. This difference arises from a different definition of the background-subtracted GOES flux.} Therefore, in the discussion section, we consider the main findings of these previous studies to infer characteristics from these events. It is remarkable to note that some of these events are as hard as X-class flares, despite being of GOES A or B class. 

    The right panel of Fig.~\ref{fig:delta&ph35-GOES} shows the photon flux at 35 keV in comparison to the background-subtracted GOES 1-8 \AA{} flux. The correlation curve deduced by \citet{2005A&A...439..737B} aligns with the RHESSI microflare statistics by \citet{2008ApJ...677..704H}. This plot reveals that hard microflares, including the event reported in \citet{2008A&A...481L..45H}, present a larger photon flux at 35 keV compared to typical events with similar SXR fluxes. Generally, hard microflares have 10 to 100 times more photon flux at 35 keV than the standard microflares of similar intensities. This suggests that an A class hard microflare typically has the photon flux at 35 keV of a standard B or C class flare.

%----------

    \subsection{HXR footpoint fluxes and distances}

    Since the reconstruction of the nonthermal footpoints of hard microflares works relatively well, we now examine their HXR fluxes and separation in the context of the classic two-footpoints flare geometry.

\begin{figure}
        \centering
        \includegraphics[width=0.48\textwidth]{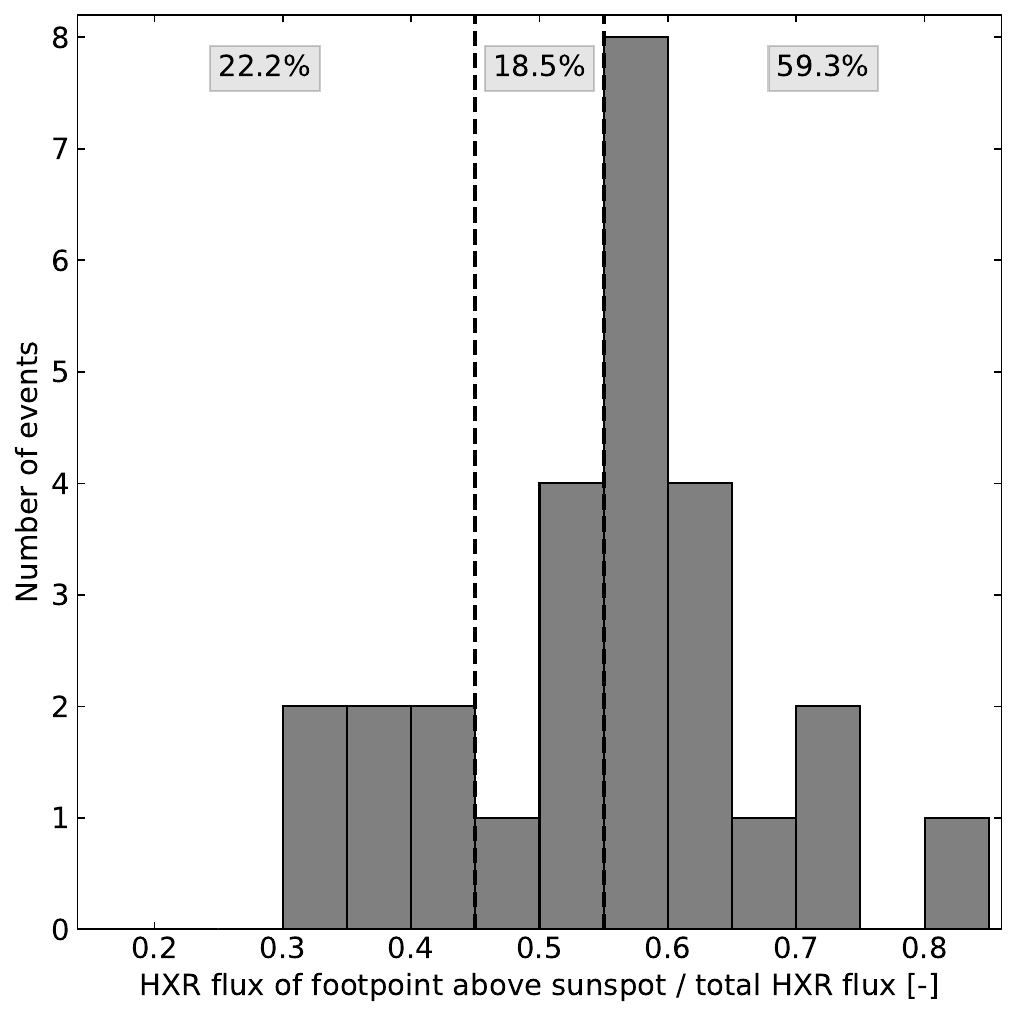}
        \caption{For the events with two HXR footpoints, this figure shows the histogram of the HXR flux coming from the footpoint above the sunspot relative to the total flux of the two HXR sources. Values above 0.5 on the horizontal axis mean that the flux is higher in the footpoint directly rooted within the sunspot, and vice versa for values below 0.5. By taking into account the error on the flux estimation, events with values between 0.45 and 0.55 (interval defined by the vertical dashed lines) are considered to have similar HXR fluxes in the footpoints. 
        Of all the events with two footpoints, $59.3\%$ have a stronger HXR flux in the footpoint rooted directly within the sunspot, $18.5\%$ have similar fluxes, and $22.2\%$ have a stronger HXR flux in the footpoint outside the sunspot.}
        \label{fig:different-fluxes}
    \end{figure}

    Figure~\ref{fig:different-fluxes} shows the histogram of the HXR flux coming from the footpoint above the sunspot relative to the total flux of the two HXR sources, for the events with two HXR sources. Among all the events with two footpoints, $59.3\%$ display a stronger HXR flux in the footpoint rooted within the sunspot. Conversely, $18.5\%$ have similar fluxes (within one sigma of the average relative error on the fluxes given by the forward fit algorithm), and another $22.2\%$ show a stronger HXR flux in the footpoint outside the sunspot.
    In addition to the events with two footpoints, there are seven events with a single HXR footpoint. By co-aligning the location of this footpoint and the corresponding thermal source with EUV and UV images, it turns out that all seven events are correlated with the ribbon in the sunspot. Therefore, for hard microflares, there is a tendency to have a stronger HXR flux at the footpoint rooted in the sunspot.
    This behavior is actually inconsistent with the expectations of electron motion and magnetic mirroring in simple magnetic loops. This issue is discussed in more detail in Sect.~\ref{subsec:asymmetric-footpoints}.
    %\afbadd{This observation is inconsistent with the expectation of electron motion in simple magnetic loops. As the magnetic moment is conserved, the electrons moving along the magnetic field lines are "mirrored" back when approaching high magnetic fields near the footpoints} \citep[e.g.,][p. 30]{2002ASSL..279.....B}. \afbadd{Thus, electrons can get trapped between mirror points, which prevents them from advancing into areas of stronger magnetic fields. The stronger the magnetic field, the higher the mirror point. The higher the mirror point, the lower the density and the less likely electrons have collisions and emit HXR bremsstrahlung. As sunspots are areas of stronger magnetic field, by assuming that the density model on both sides of the simple magnetic loop is similar, the HXR flux from the footpoint in the sunspot should be lower, which is not the case for the majority of the hard microflares. For a more detailed explanation of this topic, please refer to Sect.~\ref{subsec:asymmetric-footpoints}.}

\begin{figure}
        \centering
        \includegraphics[width=0.48\textwidth]{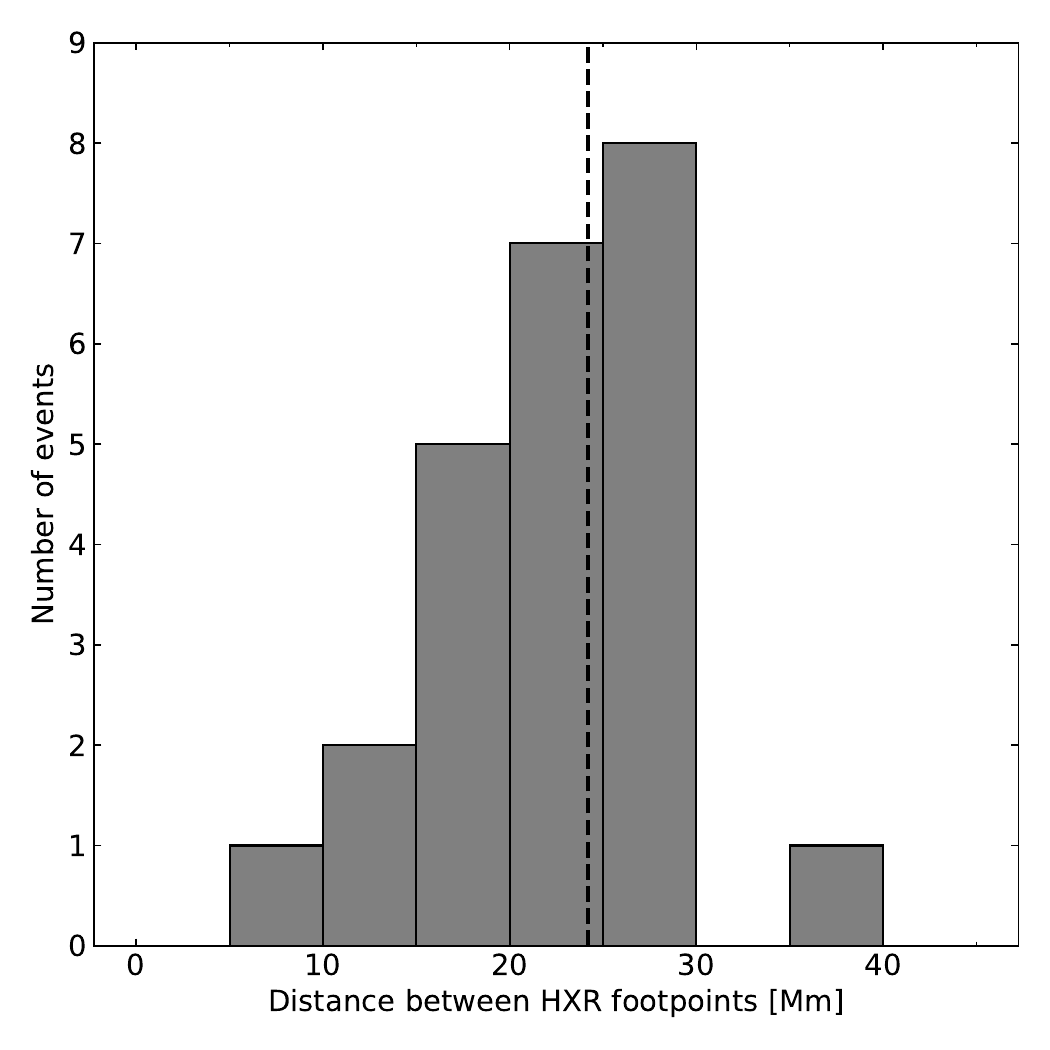}
        \caption{For the events with two HXR footpoints, we show the distribution of their distance. The vertical dashed line represents the median distance, which corresponds to approximately 24 Mm.}
        \label{fig:distance-footpoints}
    \end{figure}

    In addition to their fluxes, we estimated the distances of the two HXR sources.
    We show the results in Fig.~\ref{fig:distance-footpoints}. Most events have a separation between 15 to 30 Mm, with a median distance of about 24 Mm. These values align with \citet{2008SoPh..250...53S} for events of similar duration. %\afbcomm{Create a new paragraph after this in which we discuss the following paper (considering the mean distance we found and the flat loop): } \citet{2011SoPh..270..493C} %By assuming that the flare loop connecting the footpoints has a semi-circular geometry perpendicular to the solar surface, the altitudes obtained from this assumption are also consistent with \citet{2011SoPh..270..493C}.
    %This suggests that hard microflares are not necessarily more compact events than regular flares and other microflares with steeper spectra.

    In Sect.~\ref{sec:flare-loop-geometry}, we reported that approximately 74\% of the flare loops of hard microflares could be represented using a single-loop geometry. About one-third of them fit a semi-circle, while half with a flat loop (i.e., a semi-ellipse, where the semi-minor axis is half of the semi-major one). Therefore, we can estimate the loop-height using the separation between the footpoints. With a median separation of about 24 Mm, the loop-height would be 12 Mm for a semi-circle geometry and 6 Mm for a flat geometry. These estimates are consistent with the values found in \citet{2011SoPh..270..493C}. Hence, the loop-height of hard microflares is similar to those of standard microflares.
    
    In order to determine if there is a systematic transport effect on the electrons generating the observed HXR nonthermal flux coming from the flare footpoints, we studied the total HXR count flux from the flare footpoints versus the footpooint separation. However, no correlation between these quantities is found, suggesting that we cannot infer any potential transport effect of the electrons through the loop due to the varying length they have to travel within the loop. Other factors play a more significant role in determining the flux at the footpoints.
    %In order to determine if there is a systematic transport effect on the electrons generating the observed HXR nonthermal flux coming from the flare footpoints, we plot in Fig.~\ref{fig:total-flux-VS-distance} the total HXR count flux from the flare footpoints versus the footpooint separation. Due to the varying distance of Solar Orbiter from the Sun, we scaled the count fluxes to 1 AU. The scatter plot shows no correlation between these quantities, suggesting that we cannot infer any potential transport effect of the electrons through the loop due to the varying length they have to travel within the loop. Other factors play a more significant role in determining the flux at the footpoints.
    
    %\begin{figure}
    %    \centering
    %    \includegraphics[width=0.48\textwidth]{Figures/total-flux-VS-distance.pdf}
    %    \caption{\afbcomm{This figure can be removed.} Total HXR count flux from the flare footpoints (scaled to 1 AU) as a function of the distance between the HXR footpoints. The Spearman's correlation coefficient is -0.04 with a p-value of 0.8, indicating no correlation between these quantities.}
    %    \label{fig:total-flux-VS-distance}
    %\end{figure}
    
%----------
    
    \subsection{Line-of-sight magnetic field density}

    Given the discovery that hard microflares are directly rooted in sunspots (umbra or penumbra), we investigate the strength of the magnetic field at the flare footpoints. 
    %The two hard STIX microflares studied in \citet{2024A&A...683A..41S} revealed magnetic flux densities that are much higher than in regular cases.

\begin{figure*}
        \centering
        \includegraphics[width=\textwidth]{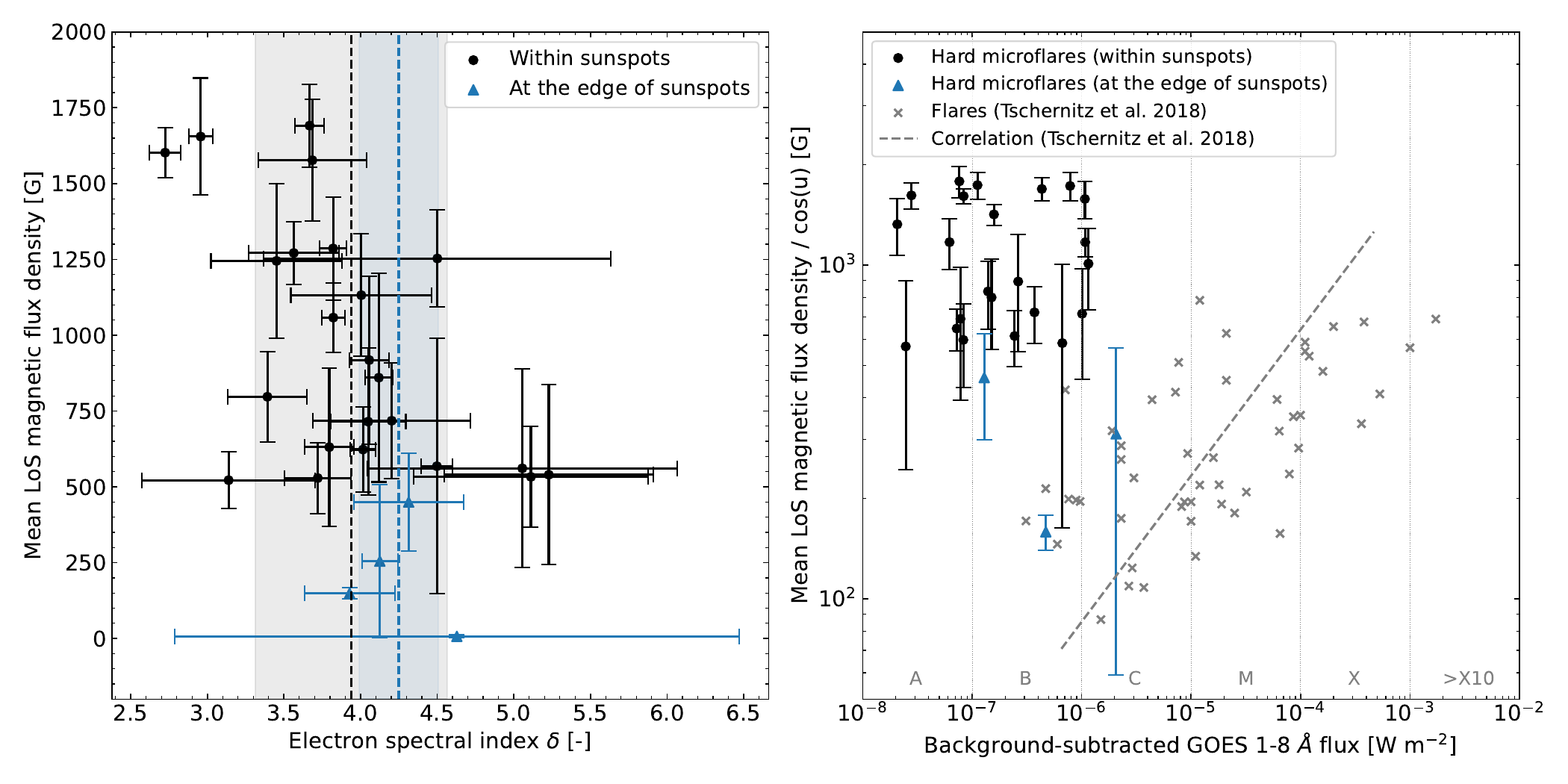}
        \caption{Analysis of the photospheric LoS magnetic flux density of the flare footpoint located within or at the edge of the sunspot. (\emph{Left}) Absolute value of the mean LoS magnetic flux density as a function of the electron spectral index. Black dots represent events with the footpoint within the sunspot, while blue triangles represent those at the edge of the sunspot. Vertical lines indicate the mean electron spectral indices, with the shaded areas representing its standard deviation. (\emph{Right}) Absolute value of the mean LoS magnetic flux density against the background-subtracted GOES 1-8 Å SXR flux. The gray crosses and the dashed gray correlation curve are derived from \citet[][Figure 9]{2018ApJ...853...41T}. To be consistent with \citet{2018ApJ...853...41T}, in this plot we multiplied the mean LoS magnetic flux density by $1 / \cos (u)$, with $u$ the angular distance from disk center of the flare location.}
        \label{fig:magnetic-field-strength}
    \end{figure*}

    The left panel of Fig.~\ref{fig:magnetic-field-strength} displays the average photospheric LoS magnetic flux density of the footpoint within the sunspot against the electron spectral index. There is a slight tendency for events within sunspots to be harder, indicated by the mean value (dashed vertical lines). This difference, however, is within the standard deviation, suggesting that a larger sample of events might provide a clearer picture. However, it is notable that events with the strongest magnetic field (> 1500 G) are among the hardest, while those with a weaker field (< 600~G) tend to show a softer spectrum.

    Because the penumbral fields are typically close to horizontal, in Appendix~\ref{sec:vec-magn-field} we performed a similar analysis as presented in the left panel of Fig.~\ref{fig:magnetic-field-strength}, but, instead of using the HMI magnetogram data, we used the HMI vector magnetic field data (see Fig.~\ref{fig:vector-magnetic-field}). The results are in agreement with the left panel of Fig.~\ref{fig:magnetic-field-strength}, with two relevant differences. First of all, the overall field strength is more elevated, because the vector magnetic field data consider the total field strength and not only the LoS component. Secondly, there is a larger sample of harder events, with respect to Fig.~\ref{fig:magnetic-field-strength}, to be associated with stronger magnetic field strengths (> 2000 G). This reflects the true nature of the strong magnetic fields in sunspots, regardless of the magnetic field orientation with respect to the LoS as it is the case for the standard magnetograms.
    
    The right panel of Fig.~\ref{fig:magnetic-field-strength} compares the magnetic flux density (multiplied by $1 / \cos (u)$, with $u$ the angular distance from disk center of the flare location) to the background-subtracted GOES 1-8 Å flux. When the flux densities associated with the hard microflares are compared to flares of various classes reported in \citet{2018ApJ...853...41T}, it becomes clear that the magnetic field is much stronger for the hard microflares. By extrapolating the expected field density from the correlation curve reported in \citet[][Figure 9]{2018ApJ...853...41T}, GOES A and B class flares should have flux densities less than 100 G. This implies that hard microflares have underlying photospheric magnetic fields that are 5 to 10 times stronger than standard microflares. In addition, for the hard microflares with the classic two-footpoints morphology, the absolute value of the mean LoS magnetic field density at the footpoint rooted within the sunspot ranges from 600 to 1800 G, whereas the outer footpoint measures from 10 to 200 G. These values align with the magnetic flux densities of two case studies reported in \citet{2024A&A...683A..41S}. For the vector magnetic field density, we obtain 1500 to 2500 G for the footpoint rooted within the sunspot, while for the outer footpoint from 100 to 400 G (see Appendix~\ref{sec:vec-magn-field}). This means that the magnetic flux density at the footpoint directly rooted within the sunspot can be about 10 times stronger than the outer footpoint.
    
    Interestingly, \citet{2018ApJ...853...41T} does not report field densities larger than 800 G, even though they considered flares up to GOES class X27. This is on one hand related to the different analysis. \citet{2018ApJ...853...41T} consider the full flare maps (over the overall flare duration) whereas here we only consider the pixels at the flare peak time. On the other hand, there is also an effect that hard microflares are anchored at photospheric field concentrations that are stronger than for regular flares.
    It is also important to note that flares have larger spatially extended ribbons compared to microflares. Therefore, averaging the field density over the ribbon area might result in a weaker field if the ribbons are spatially extended. However, this does not fail to highlight the strikingly intense magnetic flux densities observed in hard microflares, as it is also directly evidenced by their location inside the sunspots.
    
%--------------------------------------------------------------------

\section{Discussions \label{sec:discussions}}

%----------

    \subsection{Frequency of occurrence of hard microflares}

    In this subsection, we address the question of how frequently hard microflares occur on the Sun.

    The most suitable dataset for estimating the occurrence frequency of hard microflares is the one by \citet{2008ApJ...677..704H}, which includes thousands of RHESSI microflare events. According to that sample, about 1.5\% of the events fall into the hard microflare regime, defined simply as microflares (GOES class below C2, see Appendix~\ref{sec:event-selection}) with an electron spectral index lower than 5. However, the events in \citet{2008ApJ...677..704H} were detected at the peak of the thermal emission, which means that the nonthermal signal might have already significantly decreased. Therefore, this 1.5\% is an underestimation.
    Since around half of our events have a different peak time in thermal and nonthermal emission, this means that about 3\% of the events fall into the hard microflare regime, as estimated from \citet{2008ApJ...677..704H}. 
    
    As outlined in Appendix~\ref{sec:event-selection}, we detected 74 events (39 of which were observed from Earth) over a little more than 2 years of continuous STIX observations. Given that STIX observed about 20'000 flares in the same period, the number of hard microflares detected is underestimated, as 3\% of 20'000 is 600. This underestimation is likely due to the combination between the method used in this paper together with the flare list on the STIX website \citep{2023A&A...673A.142X}, and the latter is optimized for operational purposes and not science. An algorithm that relies on time profiles in the first step, rather than a pre-defined list, would be most effective for detecting all hard microflares.

    \subsection{Hard microflares in recent literature}
    
    While this paper introduces the term "hard microflares" for the first time, it is important to note that such events have been observed in the past already, particularly with such instruments as RHESSI \citep[e.g.,][]{2008A&A...481L..45H}. In this subsection, we summarize the key findings from recent literature on these events.

    %We note that these events have previously been referred to by different terms, such as "early impulsive flares" \citep[e.g.,][]{1997A&A...320..620F,2007ApJ...670..862S} or "cold flares" \citep[e.g.,][]{2016ApJ...822...71F,2018ApJ...856..111L,2023ApJ...954..122L}. However, these terms might encompass a variety of different events, not just hard microflares. We believe the term "hard microflares" describes properly these events, since the "microflare" part of the term indicates a relatively low SXR flux, while the "hard" part refers to the peculiar nature of the nonthermal spectrum.

    In their study of 27 flares, as recorded by Konus-Wind and various microwave observatories, \citet{2018ApJ...856..111L} analyzed the relationship between HXRs and microwaves. Similar to hard microflares, they found that these events are weaker, shorter, and harder in the X-ray domain compared to standard flares. In the microwave domain, the events are shorter and harder, but not weaker. Furthermore, in the microwave domain, these events often demonstrated a significantly higher peak frequency than standard flares. This high peak frequency implies strong magnetic field, which aligns with our findings, because hard microflares originate in sunspots, which are regions of strong magnetic fields. Consequently, this suggests that the events analyzed by \citet{2018ApJ...856..111L}, termed as "cold" flares\footnote{The origin of this term is due to the very low thermal response relative to nonthermal energy of accelerated particles in these events. However, the low thermal emission is not necessarily due to low temperature, but it can also be due to low emission measure at high temperature.}, are essentially hard microflares (see Fig.~\ref{fig:delta&ph35-GOES}). %\afbcomm{Maybe add a figure of one of their events?}

    \citet{2018ApJ...856..111L} report that the strong magnetic field inferred from microwaves suggests a lower altitude of the radio source, as the magnetic field decreases with altitude. This is an indication of a shorter flaring loop or lower lying. 
    Our findings suggests that the strong magnetic fields inferred from microwaves could be mainly due to the presence of the sunspot rather than the source-height. Although we find that approximately 54\% of the events (which could be approximated with a single-loop geometry) tend to exhibit a relatively flat loop (i.e., low-lying), roughly 33\% of the events are well approximated by a semi-circle geometry, indicating they are not necessarily low-lying. In addition, the estimated loop-heights agree with that of standard microflares \citep{2011SoPh..270..493C}. Therefore, the presence of the strong magnetic field due to the sunspot seems to be more important. However, the strong magnetic field inferred from microwaves can derive from a combination of both, the low-laying sources combined with the strong magnetic field of the sunspots.
    %This more compact loop implies more uniform sources, which could also explain why they observe a steeper low-frequency microwave slope in hard microflares than in standard ones. However, by comparing the footpoint distance between HXR sources, we find that hard microflares are not necessarily more compact than the standard ones. Also, in Appendix~\ref{appendix:loop-geometry}, by looking into a few examples, we showed that various flare geometries can exist, with either a flat or non-flat loop. Thus, this aspect has to be investigated in more details on case-by-case studies, including the analysis of EUV observations. These observations can provide insights into the geometry of the flare loops and provide information about their plasma and magnetic properties. 

    Let us now discuss the energy partitioning in the HXR observations of flares, which means how the total energy released during flares is distributed among different processes, such as thermal and nonthermal emission. \citet{2020A&A...644A.172W} found, after comparing results from various studies, that the thermal-nonthermal energy partition changes with flare strength. In weak flares, there appears to be a lack of energetic electrons, while in strong flares, the nonthermal energy injected is enough to account for the thermal component. However, \cite{2024ApJ...964..142A}, after analyzing several microflares that occurred in September 2021, were able to deduce the ratio between the HXR fluence (the nonthermal emission integrated over time) and the SXR flux, which is primarily thermal emission. This ratio allowed them to identify nonthermal rich flares (events with a high value of this ratio), as shown in Fig.~\ref{fig:delta&ph35-GOES}. These high ratio values suggest that the energy partitioning of these events differs from the standard trend reported by \citet{2020A&A...644A.172W}. The reason for this is not clear yet, but \cite{2024ApJ...964..142A} suggest that the plasma density in the coronal loops may play a crucial role in the thermal-nonthermal energy partition.

    %In the context of flare energetics, \citet{2020A&A...644A.130C} examined the nature of impulsive heating in weak flares. Using SDO/AIA and SDO/HMI observations, they found that a majority of coronal loops hosting hot plasma have at least one footpoint rooted in regions of interacting mixed magnetic polarity at the solar surface. They suggest that the energy release during reconnection, due to interactions between magnetic patches of opposite polarity at the solar surface, is crucial for impulsive coronal heating. A manual inspection of a few events reported in \citet{2020A&A...644A.130C}, such as the one depicted in Fig.~2 of their paper, reveals that some events are rooted near sunspots. The HXR spectrum of these events needs to be analyzed and a detailed investigation of the magnetic topology of hard microflares could shed light on whether these are the same events.
    
%----------

    \subsection{Hard HXR spectra, limited thermal response, and strong magnetic fields}

    In the following, we address two questions that are not necessarily related. The first concerns the relationship between the hard particle spectra and the limited thermal response, specifically the chromospheric evaporation. The second question is why some small flares produce such hard electron spectra.

    In response to the first question, it is important to note the two distinct types of evaporation: explosive \citep[e.g.,][]{2006ApJ...638L.117M} and gentle \citep[e.g.,][]{2006ApJ...642L.169M}. These types differ in the speed at which the heated chromospheric plasma expands into the corona. Studies have found that this speed significantly depends on the energy flux density of the beam reaching the chromospheric plasma \citep[e.g.,][]{1985ApJ...289..414F}. Simulations conducted by \citet{2015ApJ...808..177R} using the HYDRAD code \citep{2013ApJ...770...12B} indicated that lower energy electrons are more efficient at heating the atmosphere (and consequently triggering evaporation) than higher energy electrons, especially in the case of gentle evaporation. For the explosive evaporation, instead, a weaker beam flux of low-energy electron is sufficient. However, the question of why such events have a limited thermal response still remains. A potential explanation could involve the pre-flare plasma density in the loop \citep[e.g.,][]{2024ApJ...964..142A}, which might stop some low-energy electrons in the corona from reaching the chromosphere. Combined with a very flat spectrum that does not generate enough low-energy electrons to trigger significant evaporation, this could explain the limited thermal response. However, to properly answer this question, a detailed analysis of the plasma conditions of the flare loop and a spectroscopic analysis of the evaporation are needed.

    For the second question, we explore the reason why such small flares can produce hard electron spectra. Our key observational finding is that at least one of the footpoints is directly rooted in the sunspot. This suggests that the strength of the magnetic field is at the origin of the harder spectra of the accelerated electrons. However, how this compares with various particle acceleration mechanisms reported in the literature, such as DC electric field acceleration \citep[e.g.,][]{1985ApJ...293..584H,2006Natur.443..553D}, stochastic acceleration \citep[e.g.,][]{1992wapl.book.....S}, and shock acceleration \citep[e.g.,][]{1985ApJ...298..400E}, is still unclear. To address these questions in more details, we have to delve into the different models of particle acceleration and investigate the primary properties that result in hard spectra, and check whether there is a direct relationship with field strength. This intriguing question will be the focus of a future study. %\afbcomm{AFB: I left it open here, but if you have a better idea on how to conclude (or continue) this paragraph, it would be excellent!}

%----------

    \subsection{Asymmetric HXR footpoints \label{subsec:asymmetric-footpoints}}
    
    There is another interesting aspect of hard microflares, which is the asymmetric HXR emission coming from the flare footpoints, in the case of the two-footpoints morphology.

    One of the reasons for the asymmetric HXR footpoints reported in the literature is the magnetic mirroring effect \citep[e.g.,][]{1994PhDT.......335S,1996AdSpR..17d..67S,1999ApJ...517..977A}. Accelerated electrons that spirals in the direction of increasing field strength (from the loop-top downwards) encounter an opposing force, due to the Lorentz force, which may persist until its parallel velocity to the field changes sign and reflects \citep[e.g.,][p. 30]{2002ASSL..279.....B}. Hence, electrons can get trapped between mirror points, which prevents them from advancing into areas of stronger magnetic fields. The stronger the magnetic field, the higher the mirror point. The higher the mirror point, the lower the density and the less likely electrons have collisions and emit HXR bremsstrahlung. As sunspots are areas of stronger magnetic field, by assuming that the density model on both sides of the simple magnetic loop is similar, the HXR flux from the footpoint in the sunspot should be lower. However, our key finding is that approximately 78\% of hard microflares, which exhibited two HXR footpoints, have similar or even stronger flux from the footpoint rooted within the sunspot, which is inconsistent with the magnetic mirroring scenario, as in sunspots the magnetic field is much stronger relative to the surrounding regions.
    
    This inconsistency with the magnetic mirroring could have different origins. The first could be associated with the asymmetry of the reconnection point where electron acceleration happens, potentially closer to the brighter footpoint \citep[e.g.,][]{2004A&A...423..363G,2007A&A...461..285F}. %In our study, we do not conduct any investigation of the magnetic configuration leading to the hard microflares. However, it is certainly an important aspect to consider.

    The second reason is related to the asymmetric variation in plasma density along the loop \citep[e.g.,][]{2007A&A...461..285F}. In such a case, a higher HXR flux is expected at the footpoint associated with the flank of the loop with a lower density. This increases the chance for accelerated electrons to lose their energy in the chromosphere and subsequently emit HXRs from the footpoint, whereas on the other side they lose some of the energy already in the corona. In this context, it is important to investigate the density in the active region at chromospheric and coronal altitudes directly above the sunspot with respect to the surrounding areas. This is because hard microflares have one footpoint rooted in the sunspot and another outside, somewhere in the active region. However, \citet{2008A&A...481L..53T} showed that, for the active region they investigated, the density in the corona directly above the sunspot is lower compared to the surrounding area (refer to Fig. 3 of their paper). This point should also be verified in the context of hard microflares. If it were not for the strong magnetic field in the sunspot, this condition would allow electrons to deposit most of their energy (at chromospheric altitudes) at the footpoint associated with the sunspot.
    
    As previously mentioned, the mirroring effect can play a role here. It is important to note that this effect depends on the angle distribution of the accelerated electrons, hence the ratio $v/v_\perp^{\mathrm{top}}$, where $v$ represents the velocity of the electron, and $v_\perp^{\mathrm{top}}$ is its perpendicular component at the top of the loop \citep[e.g.,][]{2002ASSL..279.....B}, assuming that the reconnection point is at the top of the loop. The lower the $v_\perp^{\mathrm{top}}$, the higher the field strength that can be accessed by electrons without being mirrored. This ratio defines the pitch angle $\alpha_{\mathrm{top}}$ such as

    \begin{equation}
        \alpha_{\mathrm{top}} = \arcsin \left( \frac{v_\perp^{\mathrm{top}}}{v}  \right).
    \end{equation}

    \noindent
    Therefore, if the electron distribution is beamed, implying a narrow pitch angle distribution, electrons can penetrate the chromospheric layers, despite the strong magnetic field associated with the sunspot.

    To determine if beaming is contributing to the asymmetry and the penetration of accelerated electrons in such a strong magnetic field, one can conduct stereoscopic X-ray observations of hard microflares. Solar Orbiter/STIX provides a unique viewpoint beyond the Sun-Earth line for this purpose. As demonstrated in \citet{2024ApJ...964..145J}, this allowed us to evaluate the HXR directivity and investigate the pitch angle distribution of the accelerated electron beam.

    \subsection{Extension to regular flares}

    We discuss here the relationship between the electron spectral index and the strength of the magnetic field in the context of regular flares.

    In this context, important results are already reported in the literature and it is a combination between two different types of studies. On one hand, in this work we show how the correlation curve by \citet{2016A&A...588A.115W} aligns well with the independent statistical studies by \citet{2005A&A...439..737B} and \citet{2008ApJ...677..704H}. This implies that the intensity of the flare, on the large scale of flare intensities from A to X GOES classes, is related to the efficiency of high-energy electron acceleration. In other words, a harder spectrum is associated with more intense flares, and this applies to both flares and microflares. On the other hand, \citet{2018ApJ...853...41T} show that the more intense events correlate with stronger magnetic fields. Thus, these results suggest that a stronger magnetic field implies a harder spectrum.
    This aligns with the observation that hard microflares are associated with stronger magnetic fields (i.e., rooted in sunspots) compared to standard (and softer) microflares. These findings highlight the universality of the results.
    
%--------------------------------------------------------------------

\section{Conclusions and outlook \label{sec:conslusions}}

    In this study, we analyzed 39 hard microflares. These events have relatively low thermal emission but a notably hard spectrum at higher energies, which means that they are very efficient in accelerating high-energy electrons. Our key findings from this statistical analysis are reported in the following.

    \begin{itemize}
        \item Hard microflares have one of the footpoints directly rooted within the sunspot, which suggests that the strength of the magnetic field is at the origin of the hard spectra observed in HXRs.
        \item For the events with the classic two-footpoints morphology, the absolute value of the mean LoS magnetic field density at the footpoint rooted within the sunspot ranges from 600 to 1800 G, whereas the outer footpoint measures from 10 to 200 G. For the vector magnetic field density, we obtain 1500 to 2500 G for the footpoint rooted within the sunspot, while for the outer footpoint from 100 to 400 G. This means that the magnetic flux density at the footpoint directly rooted within the sunspot can be about 10 times stronger than the outer footpoint.
        %\item The underlying magnetic flux densities are large, with values ranging from 500 to 1800 G. This magnetic field strength is about 5 to 10 times stronger that that associated with standard microflares with softer spectra. \afbcomm{Would it here not be more important to state the flux density values in these sunspot foopoints? (like to give the mean value over al plus standard devatiation) and then maybe state the ratio to the flux densities in the other footpoint (where available).}
        \item Despite the large difference of the magnetic field at the flare footpoints, approximately 78\% of hard microflares, which exhibited two HXR footpoints, have similar or even stronger HXR flux from the footpoint rooted within the sunspot.
        \item The median footpoint separation, as measured by means of HXR observations, is approximately 24 Mm. This is consistent with events of similar GOES classes.
        \item About 74\% of the events could be approximated by a simple loop geometry, demonstrating that hard microflares typically have a relatively simple morphology. Out of these events, around 54\% exhibit a flat flare loop geometry, while roughly 33\% of the events are well approximated by a semi-circle. The rest showed a high degree of tilting. Therefore, the loop-height is estimated to range between 6 and 12 Mm.
    \end{itemize}

    Hard microflares, due to their nature, offer excellent case studies for exploring electron acceleration in flares and the thermal plasma response to nonthermal electron input. Consequently, we believe that future studies should focus on addressing three primary scientific questions: Why hard microflares exhibit a limited thermal response; what acceleration mechanism enables these small flares to generate such a hard spectrum; what is the primary factor causing the asymmetry of the HXR footpoint fluxes in these events. By addressing these questions through the study of hard microflares, we can gain a better understanding of the acceleration process and the thermal response in flares.

\begin{acknowledgements}
    The authors acknowledge Arun K. Awasthi, Iain G. Hannah, Alexandra Lysenko, and Johannes Tschernitz for providing their data, which has been included as a reference in this paper. We extend our gratitude to the anonymous referee for the constructive and supportive feedback.

    Solar Orbiter is a space mission of international collaboration between ESA and NASA, operated by ESA. The STIX instrument is an international collaboration between Switzerland, Poland, France, Czech Republic, Germany, Austria, Ireland, and Italy. IRSOL is supported by the Swiss Confederation (SEFRI), Canton Ticino, the city of Locarno and the local municipalities.
    
    AFB, SK, MZS, DFR, and HC are supported by the Swiss National Science Foundation Grant 200021L\_189180 for STIX. AFB is also supported by the Swiss National Science Foundation Grant 200020\_213147. AMV acknowledges the Austrian Science Fund (FWF): 10.55776/I4555.
\end{acknowledgements}

%--------------------------------------------------------------------

% for the bibliography, at the end
\bibliographystyle{aa} % style aa.bst
\bibliography{biblio} % your references Yourfile.bib

\begin{appendix}

%    \section{Supplementary images}

%Here I can add the figure with all UV, X-ray and manetogram images.

%\begin{figure}
%    \centering
%    \includegraphics[width=0.5\textwidth]{Figures/gong-event.pdf}
%    \caption{This is the only event without SDO data.}
%    \label{fig:enter-label}
%\end{figure}

%--------------------------------------------------------------------

\section{Automatic hard microflare selection \label{sec:event-selection}}

%Here we can include the following figures
%\begin{itemize}
%    \item time profiles and spectra of typical hard microflares
%    \item ratio as a function of time and scatter plot
%\end{itemize}
%\begin{itemize}
%    \item The following time frame has been excluded from the analysis because of the different energy binning: 2023-07-31T00:05:58 to 2023-11-05T00:01:00. However, this does not affect the statistics much because Solar Orbiter, during this time period, was facing mostly the back side of the Sun as seen from Earth and in this paper we only consider events jointly observed from Earth.
%    \item After that period, a few events where manually selected as they clearly stand out from the lightcurves and they have been added to the statistics. These are flares XXX - YYY. Their ratio reflect what has been used previously.
%\end{itemize}

%----------

    \subsection{Description of the methodology}
    
    The selection process for the hard microflares considered in this paper is detailed in the following. The complete list of the 39 selected events is provided in Tab.~\ref{tab:all-hard-microflares}.
    
    \begin{table*}[!]
        \caption{List of all the hard microflares analyzed in this paper.}
        \label{tab:all-hard-microflares}
        \begin{center}
            \fontsize{8.4}{13}\selectfont
            \begin{tabular}{|c|cccccccccc|}
                \hline %\hline
                \textbf{Event \#} & \textbf{Flare peak time}$^a$ & \multicolumn{2}{c}{\textbf{SO Coord.}$^b$} & \multicolumn{2}{c}{\textbf{Earth Coord.}$^c$} & \textbf{GOES$^d$} & \textbf{$\delta$}$^e$ [-] & \textbf{AR}$^f$ & \textbf{Category}$^g$ & \textbf{Flare loop}$^h$ \\
                 & & $x$ [$\prime\prime$] & $y$ [$\prime\prime$] & $x$ [$\prime\prime$] & $y$ [$\prime\prime$] & & & & & \\ \hline \hline
                1 &     22-Sep-2021 10:11:48      &   100   &  -830  & -422 & -555 &     B1      &     $3.7 \pm 0.2$      &   AR12871    &  within & complex \\ \hline
                2 &     22-Sep-2021 18:52:16      &     140      &   -800    & -375 & -560 &   A6  &     $5.1 \pm 0.8$      &   AR12871  &   within & flat \\ \hline
                3 &     22-Sep-2021 20:12:36      &     150      &   -800    & -400 & -530 &   B3  &     $3.8 \pm 0.2$      &  AR12871   &   within & semi-circle \\ \hline
                4 &     4-Oct-2021 06:50:46      &     -1170      &   370    & -950 & 210 &   A1  &    $4.2 \pm 0.3$       &  AR12882   &   within &  flat \\ \hline
                5 &     4-Oct-2021 08:37:02      &      -1170     &   350    & -900 & 217 &   A2  &     $4.8 \pm 0.5$      &  AR12882   &   within & flat  \\ \hline
                6 &     4-Oct-2021 09:36:58      &      -1160     &   350   & -892  & 223 &   A1  &     $3.3 \pm 0.3$      &  AR12882   &   within & flat  \\ \hline
                7 &     9-Oct-2021 00:23:04      &      10     &   340   & -240 & 150  &   <A1  &      $4.0 \pm 0.5$     &  AR12882   &   within &  tilted \\ \hline
                8 &     11-Oct-2021 19:23:50      &      780     &  330   & 345 &  165  &   A2  &     $3.0 \pm 0.1$      &  AR12882   &   within &  complex \\ \hline
                9 &     3-Feb-2022 02:14:11      &       1070    &   300   & 752 & 320  &   A3  &     $3.6 \pm 1.6$      &   AR12936  &   within &  flat \\ \hline
                10 &    3-Feb-2022 05:53:59       &      1070     &  300    & 775 & 315 &   A2   &     $3.4 \pm 0.3$      &  AR12936  &    within & flat  \\ \hline
                11 &    3-Feb-2022 12:45:24       &      1100     &  340   & 811 & 334  &   B1   &     $4.3 \pm 1.6$      &  AR12936  &    edge &  flat \\ \hline
                12 &    3-Feb-2022 20:41:59       &     1110      &  290   & 820 &  350 &   B3   &     $4.9 \pm 0.3$      &  AR12936  &    within & - \\ \hline
                13 &    30-Mar-2022 08:38:43       &      -2740     &  670    & 230 & 355 &   B1   &    $4.1 \pm 0.1$       &  AR12976  &    within &  semi-circle \\ \hline
                14 &    8-May-2022 04:20:42       &      -1190     &   -260  &  805 & -185 &   A4   &     $3.1 \pm 0.3$      &  AR13003  &    within &  complex \\ \hline
                15 &    10-Nov-2022 17:14:35       &      620     &  140   & 25 & 150  &   A4   &     $3.7 \pm 0.1$      &  AR13141  &   within &  unclear  \\ \hline
                16 &    12-Nov-2022 04:32:50       &        980   &  160    & 320 & 160 &   A4   &     $3.9 \pm 0.3$      &  AR13141  &   edge &  complex  \\ \hline
                17 &    13-Nov-2022 10:25:27       &      1230     &  250   &   576 & 192 &  A2   &     $5.1 \pm 0.7$      &  AR13141  &    within &  semi-circle \\ \hline
                18 &    24-Nov-2022 06:22:24       &      650     &  -450   & 250 & -268 &   B1   &    $4.8 \pm 0.3$       &  AR13151  &    edge & complex \\ \hline
                19 &    8-Dec-2022 07:37:56       &       530    &  -410   & 200 & -255  &   A5   &     $4.3 \pm 0.4$      &  AR13153  &    edge &  complex \\ \hline
                20 &    19-Dec-2022 08:17:20       &      -510     &  230   & -745 &  325 &   A5   &     $3.3 \pm 0.5$      &  AR13169  &   edge &  semi-circle  \\ \hline
                21 &    22-Dec-2022 14:52:04       &      240     &  230   & -84 & 330  &   A3   &     $2.7 \pm 0.1$      &  AR13169  &    within &  unclear \\ \hline
                22 &    23-Dec-2022 04:34:29       &      370     &  220    & 38 & 345 &   A8   &    $3.7 \pm 0.4$       &  AR13169  &   within &  tilted  \\ \hline
                23 &    15-Jan-2023 23:45:12       &      -410     &  240   & -730 &  345 &   A5   &     $4.4 \pm 0.3$      &  AR13192  &    within &  complex \\ \hline
                24 &    16-Jan-2023 00:29:40       &      -300     &  -350   & -660 & -216  &   A5   &     $4.5 \pm 1.1$      &  AR13190  &    within &  complex \\ \hline
                25 &    16-Jan-2023 03:44:12       &      -280     &   -350   & -637 & -210 &   B1   &     $3.8 \pm 0.1$      &  AR13190  &   within &  flat  \\ \hline
                26 &    21-Feb-2023 02:06:15       &     300      &  550   & -261 &  545 &   A6   &     $4.6 \pm 1.8$      &  AR13229  &   edge &   tilted \\ \hline
                27 &    28-Feb-2023 13:49:12       &      1000     &  610   & 380 & 535  &   A7   &     $4.2 \pm 0.5$      &  AR13234  &    within & semi-circle  \\ \hline
                28 &    11-Mar-2023 07:52:00       &      -1820     &  -230   & 325 & -280  &   <A1   &     $3.5 \pm 0.4$      &  AR13245  &    within &  flat \\ \hline
                29 &    11-Mar-2023 14:32:57       &     1120      &  -600   & 390 &  -325 &   A7   &     $4.1 \pm 0.1$      &  AR13245  &   within &  complex  \\ \hline
                30 &    11-Mar-2023 21:14:43       &      1180     &   -500   & 422 & -260 &   B1   &     $3.6 \pm 0.3$      &  AR13245  &    within &  semi-circle \\ \hline
                31 &    11-Mar-2023 22:43:00       &      1180     &  -570   & 424 & -290 &   A8   &     $4.0 \pm 0.2$      &  AR13245  &   within &  flat  \\ \hline
                32 &    12-Mar-2023 06:07:13       &     1230      &  -570   & 475 & -300  &   B2   &     $4.0 \pm 0.1$      &  AR13245  &   within &  semi-circle  \\ \hline
                33 &    12-Mar-2023 17:46:01       &      1310     &  -600    & 551 & -310 &   A2   &     $3.1 \pm 0.6$      &  AR13245  &    within &  flat \\ \hline
                34 &    3-Apr-2023 18:29:29       &      -560     &  -950   & 171 &  -310 &   A1   &     $5.1 \pm 1.0$      &  AR13270  &    within &  semi-circle \\ \hline
                35 &    18-Oct-2023 20:57:01       &      2250     &  400   & 321 & 121  &   A4   &     $5.3 \pm 0.3$      &  AR13465  &    edge &  unclear \\ \hline
                36 &    30-Oct-2023 03:48:10       &      250     &  -730   & -215 &  -346 &   B2   &     $4.5 \pm 0.1$      &  AR13474  &   within &  flat  \\ \hline
                37 &    23-Nov-2023 07:12:15       &      -450     &  170   & -525 & 230  &   C1   &     $4.1 \pm 0.1$      &  AR13492  &    edge &  complex \\ \hline
                38 &    30-Nov-2023 15:42:09       &      650     &  -510   & 400 &  -338 &   B4   &     $3.8 \pm 0.1$      &  AR13500  &    within &  unclear \\ \hline
                39 &    26-Dec-2023 23:50:53       &      900     &  -400   & 749 &  -337 &   B1   &     $4.6 \pm 0.5$      &  AR13529  &    within &  flat \\ \hline
            \end{tabular}
        \end{center}
        \vspace{0.1cm}
        
        $^a$: STIX nonthermal peak time corrected for the light travel time difference. \newline
        $^b$: Aspect corrected helioprojective coordinates in the Solar Orbiter reference frame. No additional shift is considered. \newline
        $^c$: Helioprojective coordinates in the Solar Dynamics Observatory reference frame. \newline
        $^d$: Pre-flare subtracted GOES class determined at the nonthermal peak time. \newline
        $^e$: Electron spectral index. \newline
        $^f$: NOAA active region number. \newline
        $^g$: If (at least) one of the footpoints is within or at the edge of the sunspot. No distinction is made between umbra and penumbra. \newline
        $^h$: Flare loop geometry determined from the AIA 131 \AA{} maps. There is not AIA coverage for event 12 (footpoint location determined with GONG). \newline
    \end{table*}
    
    STIX is the suitable instrument for detecting and studying hard microflares, as one of its key advantages is its constant non-solar background over the duration of the flare \citep[e.g.,][]{2021A&A...656A...4B}, a result of its deep space environment. Unlike X-ray detectors on Earth-orbiting missions, which are affected by photons produced by the interaction of high-energy particles from Earth's radiation belts and polar regions with the spacecraft, the STIX detectors maintain a stable non-solar background. Furthermore, since it orbits the Sun and is always pointing towards it, STIX doesn't have "nights," allowing for continuous observations. This uninterrupted observation and steady background make STIX ideal for detecting and analyzing relatively small events, such as microflares.
    
    Among the events considered in this paper, Fig.~\ref{fig:example-quicklook} presents the quicklook time profiles of four examples of hard microflares of different GOES classes, from A2 to C1 (background-subtracted). The peculiar behavior of these events is noteworthy, with all energy channels above 10 keV increasing impulsively around the nonthermal peak time (highlighted by the vertical black dashed line). In fact, in general, nonthermal emission is expected to be seen above 10 keV in microflares.
    
    Our primary interest is in the spectral shape of the STIX flares at the nonthermal peak, which corresponds to the time when the bulk of accelerated electrons is injected into the chromosphere, producing higher energy photons. Therefore, the first step is to determine the nonthermal peak time of all flares observed by STIX. We achieved this by starting with the STIX flare list available on the STIX data center \citep{2023A&A...673A.142X} and we then identified the nonthermal peak by detecting peaks at the most energetic channels where the signal is above the background. In our approach, we adopted a simplistic method where we only accounted for a single peak per flare, even in events where multiple peaks are present. We chose the peak with more counts in the most energetic channel, which for microflares corresponds to the hardest peak. As hard microflares typically have just one peak with a hard spectrum, this approach is well justified for sorting out such events.
    
\begin{figure*}[!h]
        \centering
        \includegraphics[width=0.7\textwidth]{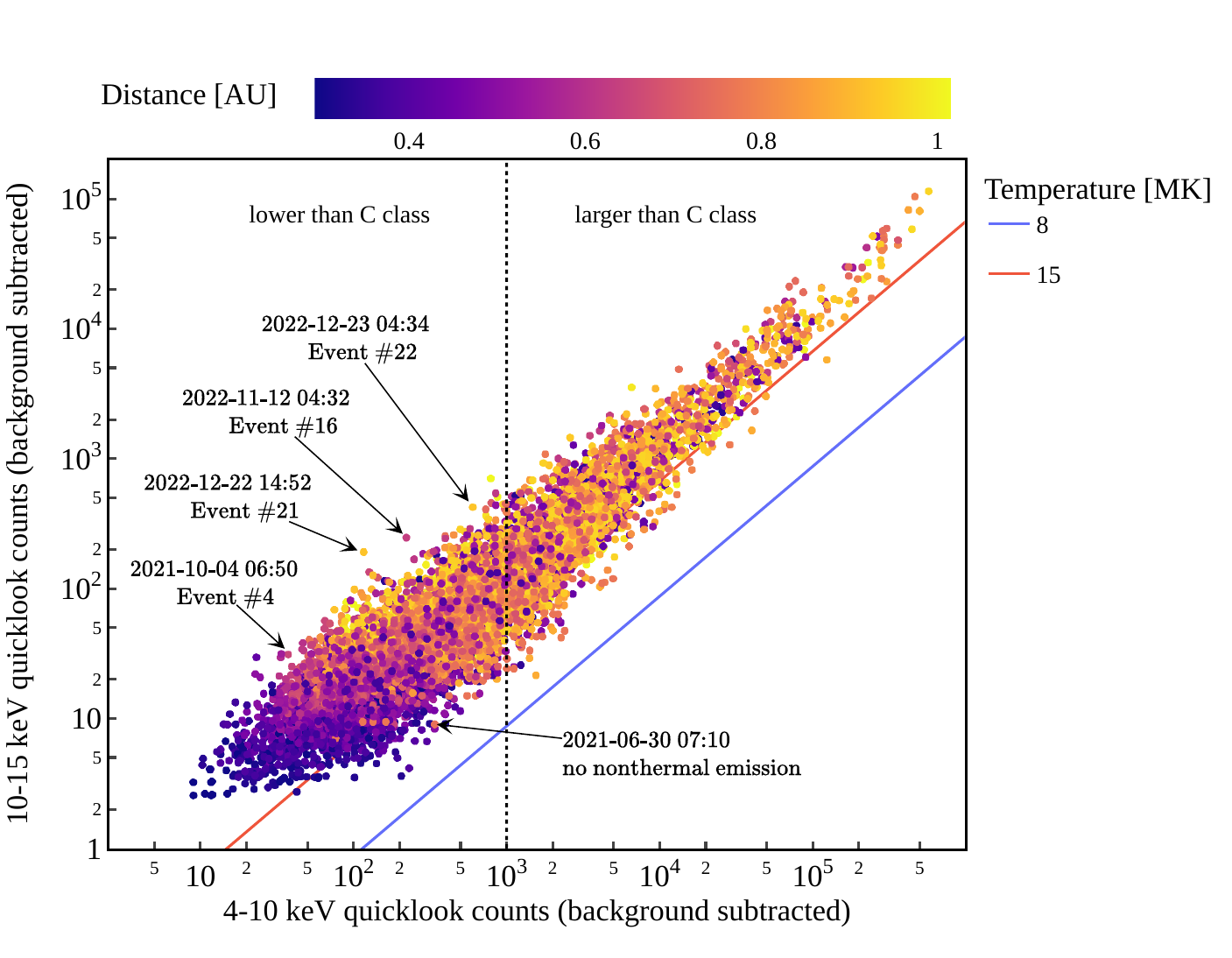}
        \includegraphics[width=0.9\textwidth]{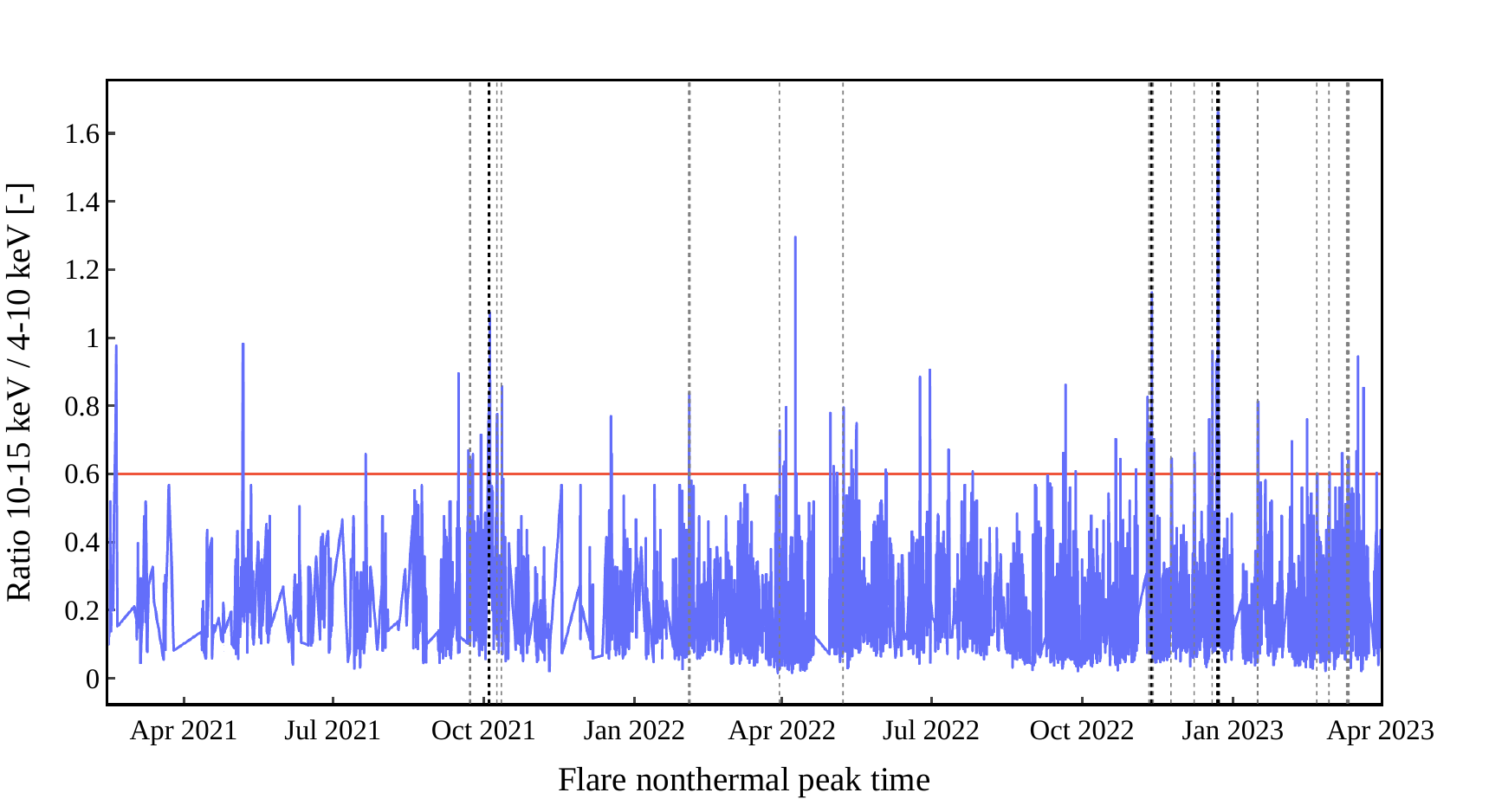}
        \caption{Summary of the event selection of the hard microflares. (\emph{Top}) Scatter plot of the 10-15 keV STIX quicklook background-subtracted counts against the 4-10 keV quicklook background-subtracted counts at the nonthermal peak, for all flares from February 2021 to April 2023. All counts have been scaled to 1 AU. The color code refers to the Solar Orbiter distance to the Sun at the time of the observation. The blue and red lines show the expected STIX quicklook counts in the case of flares with a purely isothermal component of 8 MK and 15 MK temperature, respectively. In the scatter plot, we highlighted the same events as shown in Figs.~\ref{fig:example-quicklook} and \ref{fig:example-spectra}. We can clearly observe how hard microflares cluster on the upper part of the cloud. (\emph{Bottom}) Ratio between the 10-15 keV to 4-10 keV quicklook curves as function of date, from February 2021 to April 2023. The hard microflares clearly stand out as spikes. The horizontal red line indicates the chosen threshold of 0.6 to identify hard microflares. The vertical dashed gray lines mark the events selected for this statistical analysis and the vertical dashed black lines highlight the same four hard microflares as in the scatter plot.}
        \label{fig:scatter-plot}
    \end{figure*}

    Once the nonthermal peak time of all flares is obtained, we can get the background-subtracted quicklook counts for all energy channels. To account for the varying distance of Solar Orbiter from the Sun, these peak counts are scaled to 1 AU. As previously noted, if microflares exhibit nonthermal emission, it typically occurs above approximately 10 keV \citep{2008ApJ...677..704H,2011SSRv..159..263H}. Hence, in the top panel of Fig.~\ref{fig:scatter-plot}, we plot the 10-15 keV quicklook background-subtracted counts against the 4-10 keV quicklook background-subtracted counts. As expected, a clear correlation between the two channels is evident. We also notice the increased sensitivity of STIX when Solar Orbiter is closer to the Sun. In order to select microflares, we only considered events with less than 1000 background-subtracted quicklook counts scaled at 1 AU in the 4-10 keV channel, which corresponds to events of GOES X-ray flux lower than $2 \times 10^{-6} \, \mathrm{W} \, \mathrm{m} ^{-2}$ \citep{2023A&A...673A.142X}, therefore lower than the GOES C2 class. 
    
    For microflares, assuming an isothermal model and an additional nonthermal component fitting the higher energy part of the spectrum (see Fig.~\ref{fig:example-spectra}), the vertical distribution of points in Fig. ~\ref{fig:scatter-plot} reflects the spectral index of the nonthermal spectrum. As demonstrated by the four examples in the upper part of the cloud, the harder the spectrum, the higher the counts in the 10-15 keV channel. On the other hand, lower counts in the 10-15 keV channel, as demonstrated by the SOL2021-06-30 07:10 example, tend to show soft spectra or even the absence of nonthermal emission. Consequently, hard microflares tend to cluster around the upper part of the cloud of points.
    
    Having observed the clustering of hard microflares in the scatter plot of Fig.~\ref{fig:scatter-plot}, we now need a way for their selection. In order to do this, we calculate the ratio between the background-subtracted quicklook counts of 10-15 and 4-10 keV and plot it against the date, as illustrated in the bottom panel of Fig.~\ref{fig:scatter-plot}. It is striking how hard microflares clearly stand out. By setting an arbitrary threshold value of 0.6, we can differentiate hard microflares from the standard ones. Using this value, we identified a total of 74 hard microflares between January 2021 and April 2023, with 34 being jointly observed from Earth. To the list of selected events, additional 5 events have been manually added to the list as from the STIX quicklook time profiles were clearly standing out. These five events, which range from October to December 2023, all fulfill the criteria previously described, namely less than 1000 background-subtracted STIX quicklook counts scaled at 1 AU in the 4-10 keV channel, observable from Earth, and the background-subtracted quicklook counts of 10-15 exceeding 0.6 times the 4-10 keV counts.
    
    %----------
    
    \subsection{Influence of the selection threshold \label{subsec:discussions-select-criterion}}
    
    This subsection explores the impact of the arbitrary threshold used to select hard microflares.
    
    Apart from the criterion of being observable from Earth, which reduced potential candidates by about half, the events were chosen if the ratio between the background-subtracted quicklook counts of 10-15 and 4-10 keV exceeded 0.6. The selection of this value is arbitrary, and it significantly influences the number and nature of the selected events. With a 0.6 value, about 80\% of the events are rooted within sunspots, and the rest are located at the edge of sunspots. However, Fig.~\ref{fig:magnetic-field-strength} and Fig.~\ref{fig:vector-magnetic-field} reveal a trend for the hardest flares to be located in regions of stronger magnetic field densities, hence within sunspots. The weaker and softer ones are then located at the edge of sunspots. This suggests that if we would have chosen a higher threshold value (above 0.6), we might have selected only events rooted within sunspots\footnote{The statement concerning whether all standard microflares are not rooted in sunspots has not been tested yet. This verification can be done by considering lower selection thresholds and observing if one only detects events rooted outside sunspots. This is currently the subject of an ongoing separate study.}. This is because the higher the ratio, the harder the spectrum. %In fact, by choosing a 0.7 threshold, we are only left with microflares rooted within sunspots. \afbcomm{TO CHECK.}

%%%%%%%%%%%%%%%%%%%%%%%%%%%%%%%%%%%%%%%%%%%%%%%%%%%%%%%%%%%%%%%

\section{Vector magnetic field density at the hard microflare footpoints \label{sec:vec-magn-field}}

Because in sunspots the umbral and penumbral magnetic fields have a very different inclination, in this appendix we present the analysis of the vector magnetic field density at the photosphere, as opposed to the LoS magnetic field density extracted from the magnetograms. This is achieved by analyzing the total field strength, from the hmi.B\_720s data series \citep{2014SoPh..289.3483H}, at the same footpoint locations as described in Sect.~\ref{subsec:AIAandHMI}.

\begin{figure}
    \centering
    \includegraphics[width=0.99\linewidth]{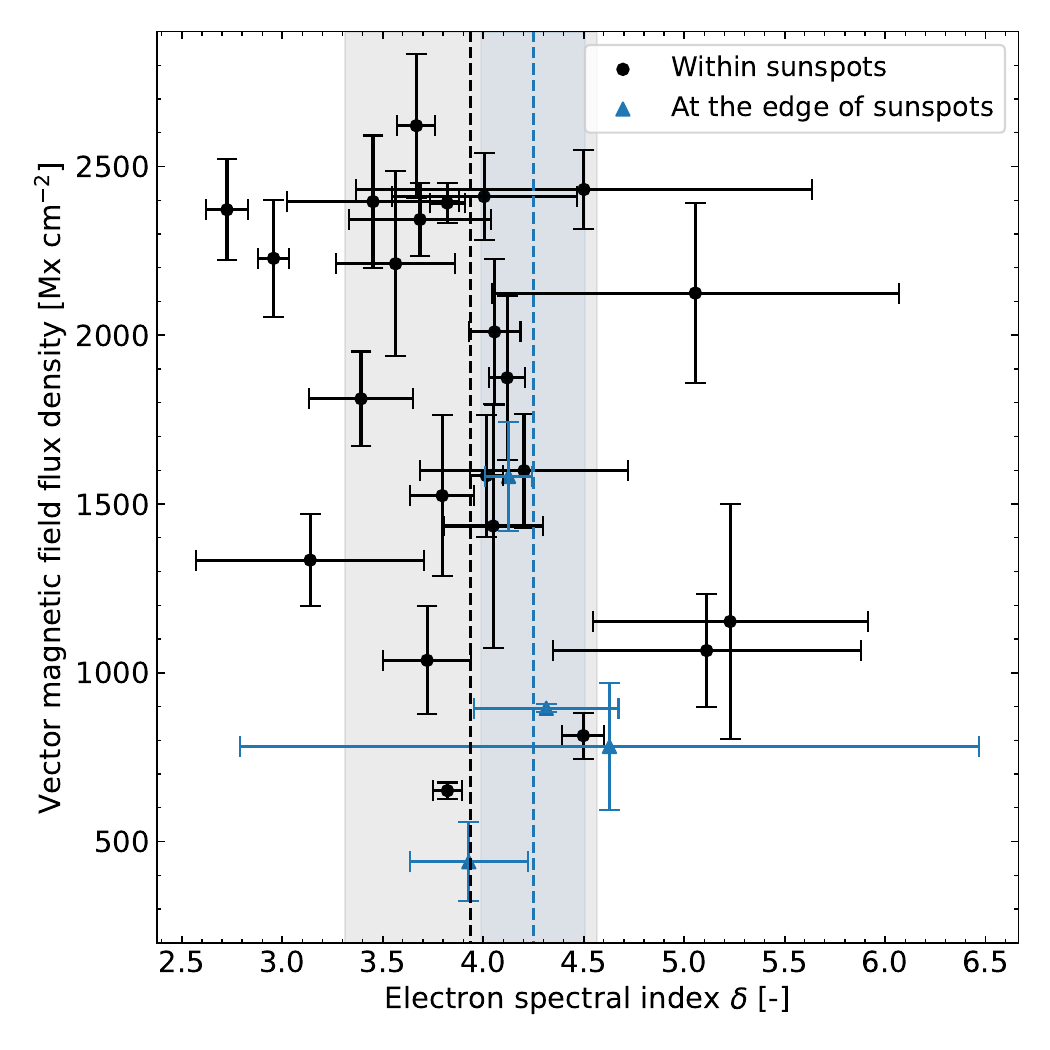}
    \caption{Analysis of the vector magnetic field density of the flare footpoint located within or at the edge of the sunspot. This figure is similar to the left panel of Fig.~\ref{fig:magnetic-field-strength}, with the difference that here we consider vector magnetic field data. We note that $1\,\mathrm{Mx}\,\mathrm{cm}^{-2} = 1\,\mathrm{G}$.}
    \label{fig:vector-magnetic-field}
\end{figure}

In Fig.~\ref{fig:vector-magnetic-field}, the vector magnetic field flux density is shown against the electron spectral index $\delta$. The trend is similar and the results are consistent with what we reported in Fig.~\ref{fig:magnetic-field-strength}. However, some interesting differences are worth to be highlighted. First of all, the overall field strength is more elevated in this case, because we are considering the total field strength and not only the LoS component. Secondly, we note a larger sample of harder events, with respect to Fig.~\ref{fig:magnetic-field-strength}, to be associated with stronger magnetic field strengths (> 2000 G). This reflects the true nature of the strong magnetic fields in sunspots, regardless of the magnetic field orientation compared to the LoS, as it is the case for the standard magnetograms.

In the case of the hard microflares with the classic two-footpoints morphology, the mean field strength at the footpoint rooted within the sunspot ranges from 1500 to 2500 G, whereas the outer footpoint measures from 100 to 400 G. This difference is consistent with the results obtained in the context of the standard magnetogram data, as the field strength at the footpoint rooted within the sunspot can be about 10 times stronger than that of the outer footpoint.

\end{appendix}

\end{document}